\documentclass[12pt]{article}
\usepackage{graphicx,psfrag,epsf}
\usepackage{enumerate}
\usepackage{natbib}
\usepackage{url} %
\usepackage{caption, subcaption, float}
\usepackage[margin=1in]{geometry}

\usepackage{amsmath,amsthm,amssymb,graphicx,mathtools,enumerate, textcomp, bbm, hyperref, color, booktabs}
\hypersetup{colorlinks,linkcolor={blue},citecolor={blue},urlcolor={red}}  
\usepackage[title]{appendix}
\usepackage[linesnumbered,ruled,lined]{algorithm2e}
\usepackage{xcolor}
\usepackage{comment}
\usepackage{dsfont}
\usepackage{bm}
\usepackage{changepage}

\renewcommand*\theassumption{(A\arabic{assumption})}
\newcounter{subassumption}[assumption]

\makeatletter
\renewcommand{\p@subassumption}{\theassumption}%
\makeatother

\numberwithin{equation}{section}

\numberwithin{equation}{section}

\newcommand{\set}[1]{\left\{#1\right\}}

\graphicspath{ {./} }

\begin{document}

\def\spacingset#1{\renewcommand{\baselinestretch}%
{#1}\small\normalsize} \spacingset{1}

\title{Quantifying Time-Varying Physical Activity Intervention Effects via Functional Regression}

\author{
Nidhi Pai$^{1}$\thanks{Email: pai00032@umn.edu}
\and
Yu Lu$^{2}$
\and
Kristin A. Linn$^{3}$
\and
Erjia Cui$^{1}$
}

\date{
\small
$^{1}$Division of Biostatistics and Health Data Science, University of Minnesota\\
$^{2}$Department of Biostatistics, Johns Hopkins University\\
$^{3}$Department of Biostatistics, Epidemiology, and Informatics, University of Pennsylvania
}

\maketitle

\begin{abstract}
Physical activity (PA) intervention studies often collect repeated intensity measurements over long observation periods.
Quantifying the variation in intervention effects over the study period is critical to evaluating and improving intervention strategies, yet many analyses reduce PA data into scalar summary measures, resulting in limited insights.
We propose a functional regression framework, which captures time-varying intervention effects by modeling the entire PA trajectory as a functional observation.
From both methodological and practical perspectives, we demonstrate the advantages of function-on-scalar regression (FoSR) over the traditional two-step approach of applying functional principal components analysis (FPCA) followed by regressing scores on covariates.
The FoSR is further extended to a function-on-function regression (FoFR) for studying the association of PA across time periods.
Methods are applied to daily step counts from the Social incentives to Encourage Physical Activity and Understand Predictors (STEP UP) study, revealing distinct and highly interpretable time-varying effects of three intervention strategies on PA and differences in their sustainability.
Our case study highlights the feasibility of functional data analysis techniques for uncovering novel insights in intervention studies with high-dimensional endpoints.
\end{abstract}


\newpage
\baselineskip=24pt

\section{Introduction}
\label{sec:intro}
Physical activity (PA) is a key modifiable factor in preventing chronic diseases like hypertension and diabetes \citep{physical_acitivity_warburton2006}. With only one in four U.S. adults meeting the recommended PA guidelines \citep{cdc_physical_activity}, there is a critical need for interventions that promote and sustain PA, as even small increases can be beneficial \citep{physical_activity_guidelines}.
Wearable devices have emerged as a reliable method for objectively measuring PA intensity over extended periods. Given their accuracy and affordability, these devices are increasingly deployed in PA intervention studies \citep{be_fit_patel2017, pa_intervention_zhang2016, pregnancy_intervention_lewey2022}. However, analyzing the high-dimensional data collected by these devices is arduous.

\begin{figure}
    \centering
    \includegraphics[width=\linewidth]{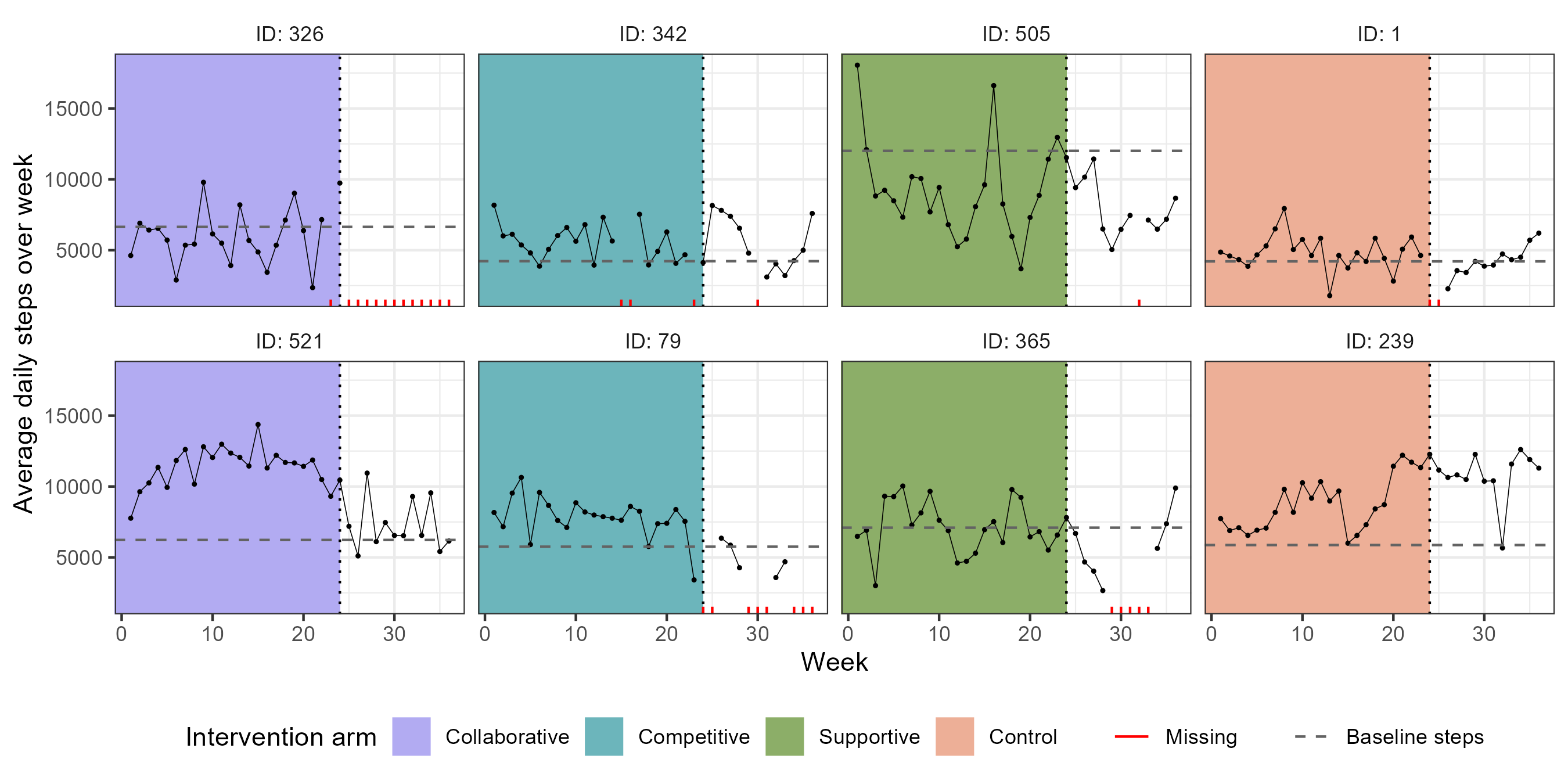}
    \caption{Selected participants' daily average step counts from STEP UP. Two participants from each arm are included.
    Data is averaged for each week (Section~\ref{subsec:preprocessing}), and if a participant has less than 3 observations in that week, the week is considered missing.
    The intervention ran to week 24 (colored region), and follow-up ran for 12 weeks after.}
    \label{fig:step_up_data_plot}
\end{figure}

Our motivating example is Social incentives to Encourage Physical Activity and Understand Predictors (STEP UP), a randomized clinical trial aimed at increasing PA through gamified behavioral interventions \citep{step_up_primary}.
The study consisted of 602 individuals randomized to four arms: collaborative, competitive, supportive, or control.
Each participant wore a device to track their daily steps for the 24-week intervention and 12-week follow-up periods.
Figure~\ref{fig:step_up_data_plot} displays average daily steps by week over the study period for eight individuals, two from each study arm.
The gray dashed line denotes their pre-intervention baseline daily steps, and red ticks denote weeks with fewer than 3 days of recorded steps.
Close inspection of Figure~\ref{fig:step_up_data_plot} reveals substantial within- and between-arm heterogeneity, as well as distinct activity patterns between the intervention and follow-up periods.
Determining how to appropriately account for such variability in modeling poses a significant statistical challenge.

\begin{figure}
    \centering
    \includegraphics[width=1\linewidth]{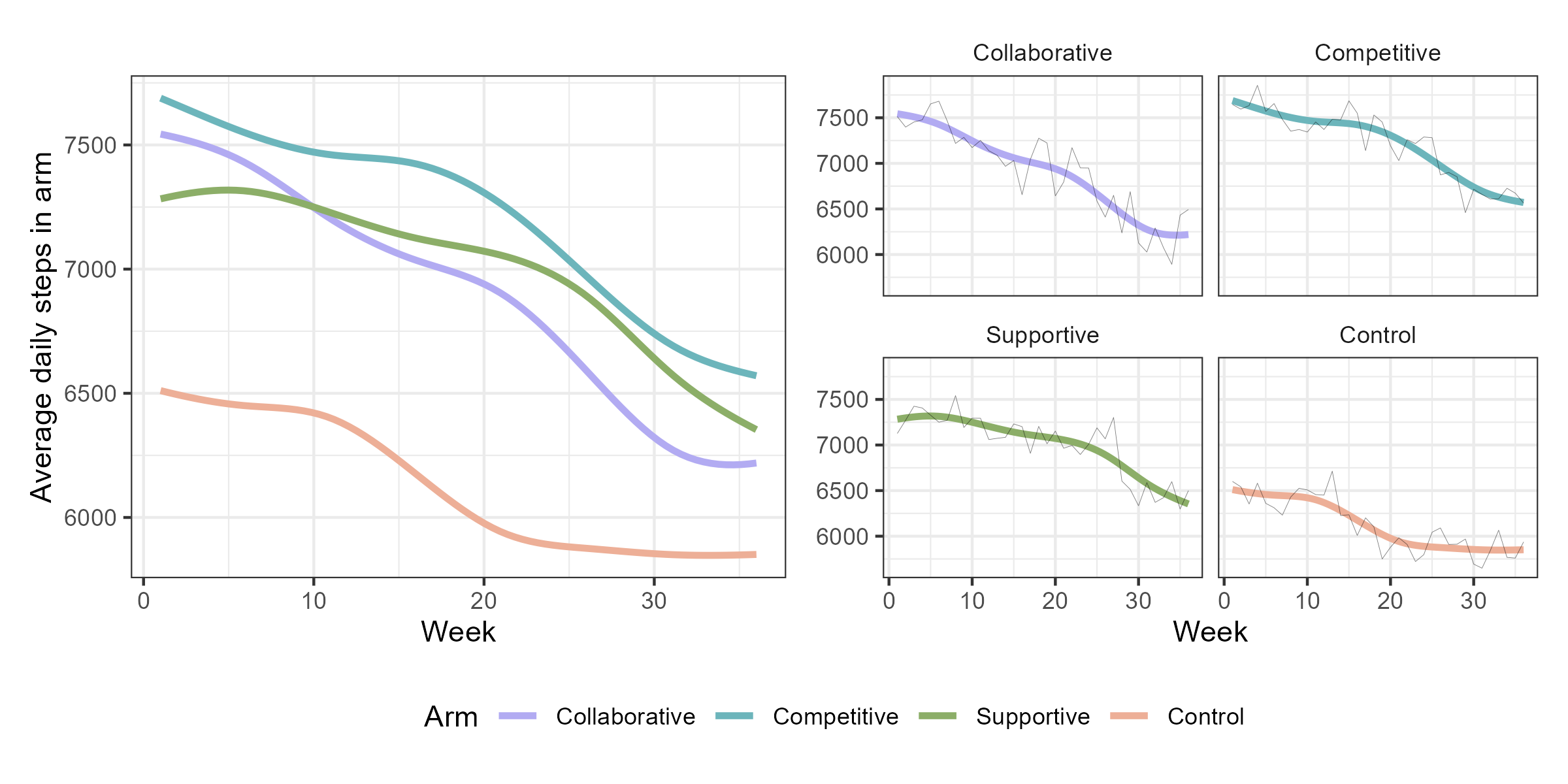}
    \caption{Smoothed average daily steps by arm (thick lines). The thinner lines in the right panels are the average daily steps by arm before smoothing.}
    \label{fig:smoothed_avg_plot}
\end{figure}

Previous analyses of STEP UP focused on a single, time-independent effect for each intervention over the entire intervention (or follow-up) period.
Using generalized linear mixed models, \cite{step_up_primary} concluded that all three intervention arms had a significantly greater mean increase in steps from baseline compared with the control arm during the intervention period, while only the competitive arm sustained significantly greater PA in the follow-up period.
However, collapsing each participant's step count trajectory into summary measures results in the loss of important temporal information.
As an example, the left panel of Figure \ref{fig:smoothed_avg_plot} shows the smoothed average daily steps by arm, where each color represents one arm.
The average steps before smoothing are shown in light gray in separate panels on the right.
Despite a decreasing trend across all groups, the nonparallel lines between intervention and control groups indicate that the intervention effects may vary across time points.
Previous analyses did not account for time-varying effects, limiting their attention to between-arm differences in monthly average step counts without covariate adjustment or inference.
Furthermore, although the step counts were significantly higher in the intervention groups, it remains unclear how long these effects last after the intervention.

To answer these critical questions, we introduce a functional regression framework, which models
each individual's PA trajectory as a functional observation. Functional data analysis (FDA) \citep{ramsay_silverman_fda, morris2015review, fda_with_r_crainiceanu} has become increasingly popular due to the collection of high-dimensional data in multiple fields.
Compared with traditional methods, FDA methods offer two important advantages for STEP UP: (1) they enable the quantification of changes in PA over the entire time frame by avoiding excessive data compression; (2) they naturally accommodate missing data or irregular observations by assuming an underlying smooth function for the observed trajectory.
This is especially important for our case study, where missing data may arise from participants disengaging from the study or forgetting to wear or replace the battery in their devices.

Specifically, our framework consists of two parts.
In the first part, we introduce a function-on-scalar regression (FoSR) model \citep{reiss2010_fosr} to study the effectiveness of PA interventions by regressing PA trajectories on study arms and scalar covariates.
This approach, although less commonly used in applied settings than the traditional two-step ``functional principal components analysis + regression" method (e.g., \cite{kringle2023fpcaobesity, xu2019mfpca_pa}), offers several key advantages. Using the STEP UP analysis, we demonstrate that FoSR improves upon the two-step model by mitigating several of its limitations and is less sensitive to parameter selections. In the second part, we introduce a function-on-function regression (FoFR) \citep{ivanescu2015penalized_fofr, scheipl2015famm} model to predict follow-up PA trajectory based on baseline steps, arm, and intervention-period PA.
This question, though not investigated in previous STEP UP analyses, is crucial for understanding the sustainability of PA interventions.
While our framework is demonstrated through STEP UP, the methods are applicable to other intervention studies with high-dimensional endpoints (e.g., repeated measures of pain).
Reproducible software is provided to facilitate methods implementation in different settings.

The rest of the paper is organized as follows. We describe the STEP UP data in Section~\ref{sec:data_description}. Section~\ref{sec:methods} introduces our functional regression framework. The application results are shown in Section~\ref{sec:results}.
Section~\ref{sec:sim_study} contains a simulation study.
We conclude with discussions in Section~\ref{sec:discussion}.

\section{Data Description}
\label{sec:data_description}

The Social incentives to Encourage Physical Activity and Understand Predictors (STEP UP) study was a randomized clinical trial evaluating the effect of behavioral interventions with gamification on increasing PA.
Gamification refers to the incorporation of game elements such as points and levels in non-game settings.
A total of 602 Deloitte Consulting employees in the United States, each with a self-reported BMI of 25 or higher, were enrolled between February 2018 and March 2019.
The study consisted of a 2-week run-in period, followed by a 24-week intervention period and a 12-week follow-up period.
The daily step count of each participant was tracked throughout the study using a wrist-worn device (Withings/Nokia Activité Steel).
Each participant had a baseline step count, estimated as the average daily steps from the run-in period, and selected a daily step goal.
For more details on the design, see \cite{step_up_design} and \cite{step_up_primary}.

\subsection{Study Arms and Covariates}
\label{subsec:covariates}

STEP UP included three intervention arms and a control arm, each with around 150 participants.
Participants were randomized at enrollment into one arm by stratifying on baseline steps ($<$5000, 5001-7500, or $>$7500 steps per day) in blocks of four groups with three participants each.
Participants in the control arm received no interventions besides a standard data dashboard from the device.
Participants in the competitive arm were placed into groups of three and shown a leaderboard displaying cumulative points and current level.
In the supportive arm, participants selected a friend or family member who was given updates on their performance and instructed to provide encouragement.
In the collaborative arm, participants were placed into groups of three, and one member was randomly selected to represent the team each day.
Each participant started with 70 points at the beginning of each week and retained 10 points for every day they reached their step goal.
In the competitive and supportive arms, retaining points was based on the participant reaching their personal step goal, whereas for the collaborative arm, points for all group members were based on whether the selected member reached their goal.
Participants advanced a level if they had at least 40 points at the end of the week.
In the follow-up period, all participants retained their wearable devices and data dashboard, but the games were discontinued.

A wide variety of demographic and health information, as well as validated survey instruments, were collected at baseline.
We consider the variables included in \cite{step_up_latent}. The baseline covariates fall into five categories:
\begin{enumerate}
    \item Demographics: Age, gender, and previous use of wearable devices (yes/no).
    \item Health behaviors: Baseline step count and sleep quality \citep{pittsburgh_sleep}.
    \item Personality: Big Five traits \citep{big_five_1999} and grit \citep{grit_short_scale}.
    \item Behavioral perceptions: The sticking-to-it component of exercise self-efficacy (ESE) \citep{ese_self_efficacy} and risk-taking specific to health/safety and social behavior \citep{dospert_risk}.
    \item Social support: Social support in a medical context \citep{mos_social_support}.
\end{enumerate}
The start date for each participant varied between February 12, 2018, and July 9, 2018; participants in the competitive or collaborative arms started with others in their cohort of three.
In the analysis, we include ``start day", defined as the number of days between the start date of the participant and start date of the study (February 12, 2018).

\subsection{Preprocessing}
\label{subsec:preprocessing}

Following \cite{step_up_primary}, daily step counts less than 1000 were treated as missing.
Given that the intervention was conducted on a weekly basis, daily steps were averaged for each week, labeling a week as missing if the participant had fewer than three days of recorded step counts.
Missing data increased as the study progressed, which may be due to participants forgetting to wear the device, not replacing the battery, or disengaging from the study.
After preprocessing, 127 out of 602 participants had data for each week in the 36-week study period.
Supplementary Figure \ref{fig:s_miss_by_treat} shows the proportion of participants in each arm with step counts recorded over days in the study.
The competitive arm has more observations than the other arms, especially around the end of the intervention period.
In our framework, step counts are treated as irregularly spaced observations from smooth underlying curves, rather than data on a fixed grid with missing values, so the methods can handle some missing data.
Additionally, following the literature \citep{varma2018total, cui2021additive}, the average step counts were further log-transformed to reduce right-skewness as shown in Supplementary Figure \ref{fig:s_log_transform_qq}.
Sensitivity analyses to preprocessing are reported in the Supplementary Section \ref{supp:sensitivity}. There, we repeat the main FoSR analysis with the following modifications: (1) using daily instead of weekly data, (2) labeling a weekly measure as missing only if there were 0 days of data, (3) imputing daily steps with FPCA before weekly averaging, (4) omitting the log transformation, and (5) using different covariates.

\section{Methods}
\label{sec:methods}

\subsection{Function-on-Scalar Regression}
\label{subsec:methods_fosr}

A function-on-scalar regression (FoSR) model analyzes the association between scalar predictors and a functional response.
Let $\{W_i(t) : t \in \mathcal{T}\}$ be the functional response of the $i$th subject, where $i = 1, ..., n$, and $\mathcal{T}$ is the functional domain.
In STEP UP, $W_i(t)$ is the observed step count for individual $i$ at week $t$ of the study.
We model the intervention and follow-up period separately because they may be incomparable, so $\mathcal{T} = [1, 24]$ for the intervention period and $\mathcal{T} = [25, 36]$ for the follow-up period.
Denote by $X_{im}$ the $m$th covariate of the $i$th subject where $m = 1, ..., M$. We assume the first $Q$ variables have varying effects $\beta_m(t)$  over $\mathcal{T}$ and the remaining have time-invariant effects $c_m$.
The model is
\begin{equation}
    W_i(t) = \beta_0(t) +  \sum_{m = 1}^{Q} X_{im} \beta_m(t)
    + \sum_{m=Q + 1}^M X_{im} c_m
    + Z_i(t) + \epsilon_i(t)\;,
    \label{eq:fosr_funccoef}
\end{equation}
where $\epsilon_i(t) \sim N(0, \sigma^2_{\epsilon})$ and are independent.
$Z_i(t)$ is a subject-level functional random effect capturing the residual correlation across $\mathcal{T}$.
This is a necessary component in FoSR and is assumed to be a smooth function \citep{morris2015review}.
We estimate the functional fixed effect $\beta_m(t)$ with a basis expansion as $\beta_m(t) = \sum_{l=1}^L b_{lm} B_l(t)$ for $m = 0, 1, \cdots, Q$, where $\set{B_l(t)}$ are known, pre-specified basis functions.
We use P-splines \citep{eilers_marx1996_psplines}.
To expand $Z_i(t)$, we use the estimated eigenfunctions $\phi_k(t)$ of $W_i(t)$, but one could also use a spline basis or eigenfunctions of the residuals from the model without random effects \citep{fda_with_r_crainiceanu}.
Specifically, $Z_i(t) = \sum_{k = 1}^K \zeta_{ik} \phi_k(t)$, where $\zeta_{ik}$ are independent random variables following $N(0, \sigma^2_{\zeta_k})$. After these expansions, the model becomes
\begin{equation}
    W_i(t) = \sum_{l=1}^L b_{l0} B_l(t) + \sum_{m = 1}^{Q} X_{im} \sum_{l=1}^L b_{lm} B_l(t)
    + \sum_{m=Q + 1}^M X_{im} c_m
    + \sum_{k = 1}^K \zeta_{ik} \phi_k(t) + \epsilon_i(t)\;.
    \label{eq:fosr_eq}
\end{equation}
To induce smoothness and prevent overfitting, each functional coefficient (including the intercept) is penalized by $P(\boldsymbol{b}_m) = \lambda ||D \boldsymbol{b}_m||^2$, where $\boldsymbol{b}_m = (b_{1m}, ..., b_{Lm})^T$ is a vector of basis coefficients, $D$ is a second-order difference matrix, and $\lambda$ is a parameter controlling the degree of smoothness.
This is equivalent to treating all $L (Q+1)$ basis coefficients $b_{lm}$ as random effects
\citep{semiparametric_regression_ruppert2003}.
Since $B_l(t)$ and $\phi_k(t)$ are pre-specified and known, model \eqref{eq:fosr_eq} reduces to a linear mixed effects model with two forms of random effects: $\zeta_{ik}$ are random effects from subject-specific terms, and $b_{lm}$ are random effects due to the penalty. The FoSR model naturally accommodates irregular observations, since the model remains estimable as long as $W_i(t)$ is not completely missing for subject $i$.
See Supplementary Section \ref{supp:fosr_model} for further discussion on the functional random effects and penalized model fitting.

There are several software options to fit FoSR models, including the function \texttt{pffr()} in the \texttt{R}  package \texttt{refund} \citep{refundR}, which is a user-friendly wrapper on \texttt{gam()} in \texttt{mgcv} \citep{mgcvR}.
Here we directly use \texttt{mgcv::gam()} due to its flexibility to model subject-specific random effects using an FPCA basis.
To fit the model with \texttt{mgcv::gam()}, the data needs to be organized in long format; see the Supplementary Materials for implementation details.
Functional data with $n$ individuals and $p$ time points has $O(np)$ rows in long format, making even modest datasets computationally challenging.
For example, the STEP UP dataset contains 13,920 rows in long format.
To improve computational speed, we use \verb|mgcv::bam()|, a modification of \verb|gam()| designed for large datasets.

\subsection{Limitations of FPCA + regression}
\label{subsec:methods_fpca}

A common approach to modeling functional responses and scalar predictors is to first apply functional principal components analysis (FPCA) to the functional observations and then regress the leading principal component (PC) scores on covariates, hereafter referred to as the ``FPCA + regression" approach.
Using the same notation introduced in Section~\ref{subsec:methods_fosr}, by Kosambi--Karhunen--Lo\`{e}ve theorem \citep{karhunen1947under} the FPCA decomposition of $W_i(t)$ is
\begin{equation}
    W_i(t) = \mu(t) + \sum_{k = 1}^\infty \xi_{ik} \phi_k(t) + \epsilon_i(t)\;,
    \label{eq:fpca_eq}
\end{equation}
where $\mu(t)$ is the population average curve, $\phi_k(t)$ are eigenfunctions, $\xi_{ik}$ are independent scores, and $\epsilon_i(t) \sim N(0, \sigma_{\epsilon}^2)$ are independent errors.
For dimension reduction, the expansion is truncated to the first $K$ eigenfunctions, where $K$ is often selected by the percentage variability explained (PVE) \citep{face_xiao2016}.
To conduct FPCA, we used the \verb|refund::fpca.face()| function, which is computationally fast by incorporating the FACE algorithm \citep{face_xiao2016} and handles missing data by alternating between smoothing the covariance matrix and predicting missing values.
The second step regresses each score on the baseline variables. Specifically, the regression model for the $k$th score is
\begin{equation}
    \xi_{ik} = \gamma_{k0} + \sum_{m = 1}^{M} X_{im} \gamma_{km} + e_{ik}\;,
    \label{eq:reg_eq}
\end{equation}
where $e_{ik} \sim N(0, \sigma^2_{ek})$ are independent.

To understand the limitations of this two-step approach compared to FoSR, substituting $\xi_{ik}$ in equation \eqref{eq:reg_eq} into the FPCA model \eqref{eq:fpca_eq} gives
\begin{equation}
\begin{aligned}
    W_i(t) &= \mu(t) + \sum_{k=1}^K \gamma_{k0} \phi_k(t)
    + \sum_{m=1}^{M} X_{im} \sum_{k = 1}^{K} \gamma_{km} \phi_k(t)  + \sum_{k=1}^K e_{ik} \phi_k(t) + \epsilon_i(t) \\
    &= \gamma_0(t) + \sum_{m=1}^{M} X_{im} \gamma_m(t)  + \sum_{k=1}^K e_{ik} \phi_k(t) + \epsilon_i(t)\;,
    \label{eq:fpca_reg_eq}
\end{aligned}
\end{equation}
where functional coefficients are defined as $\gamma_0(t) := \mu(t) + \sum_{k=1}^K \gamma_{k0} \phi_k(t)$, and $\gamma_m(t) := \sum_{k = 1}^K \gamma_{km} \phi_k(t)$ for $m = 1, ..., M$.
By introducing this notation, the model resembles the FoSR model \eqref{eq:fosr_eq}, as both equations \eqref{eq:fosr_eq} and \eqref{eq:fpca_reg_eq} contain a functional intercept, functional coefficients, functional random effects, and independent errors.

However, a closer inspection reveals several key limitations of the two-step approach.
First, the coefficients in FPCA + regression, $\gamma_m(t)$, are more constrained because they must be in the span of the first $K$ eigenfunctions from FPCA, where $K$ is typically kept small for interpretability.
In contrast, FoSR offers greater flexibility by using a large number $L$ of basis functions $\set{B_l(\cdot)}$ and accommodating a wide range of basis types.
Second, the shape of the coefficients in FPCA + regression are very sensitive to the choice of $K$ and the estimated eigenfunctions, as no constraints are imposed.
However, in FoSR, since the coefficients are regularized by the penalty, they are not heavily dependent on the basis functions \citep{goldsmith_penalized_fr_2011}. As we will see, this issue is exactly what arises in our application.
Finally, in the two-step approach, the error from estimating scores $\xi_{ik}$ in FPCA is propagated into the regression. Moreover, unlike fixed effects parameters, this uncertainty in score predictions does not decrease as sample size increases, making the analysis results extremely sensitive to outliers. The FoSR model does not have such a problem because it is a joint modeling.

In summary, the FoSR model addresses several key issues related to flexibility, estimation stability, and error propagation inherent in the traditional two-step FPCA + regression approach, making it a more suitable choice for STEP UP applications.

\subsection{Function-on-Function Regression}
\label{subsec:methods_fofr}

The FoSR model introduced in Section~\ref{subsec:methods_fosr} provides a flexible solution to quantify the time-varying effectiveness of interventions on PA.
To understand the sustainability of the interventions, we model the association between steps in the intervention period and steps in the follow-up period by extending the FoSR to a function-on-function regression (FoFR) model.
Let $W_i(t)$ be the step count for individual $i$ at time $t \in \mathcal{T} = [1, 24]$, the intervention period, and let $V_i(u)$ be the step count at time $u \in \mathcal{U} = [25,36]$, the follow-up period.
Denote by $A_{i1}, ..., A_{iP}$ the indicator variables for each study arm, where $P=4$ in STEP UP. In addition, assume we observe scalar variables $X_{i1}, ..., X_{iM}$, where the first $Q$ variables have effects varying over $\mathcal{U}$ and the rest have time-invariant effects. The model is
\begin{equation}
    V_i(u) = \mu(u) + \sum_{p = 1}^P A_{ip} \int_\mathcal{T}  W_i(t) \beta_p (t, u) dt + \sum_{m = 1}^Q X_{im}\beta_m(u)
    + \sum_{m=Q + 1}^M X_{im} c_m
    + Z_i(u) + \epsilon_i(u)\;,
    \label{eq:fofr}
\end{equation}
where $\epsilon_i(u) \sim N(0, \sigma^2_\epsilon)$.
Similar to FoSR, $Z_i(u)$ is a subject-level functional random effect, and $\beta_m(u)$ quantifies the time-varying effect of a scalar covariate $X_{im}$ on $V_i(u)$.
Because the effects of different strategies may vary, we allow the effects of the functional predictor $W_i(t)$ to differ across arms by assuming a different $\beta_p(t, u)$ for each arm $p$.
To estimate the coefficients $\beta_p(t, u)$, we use bivariate splines, specifically, a tensor product of splines such that $\beta_p(t, u) = \sum_{k_1 = 1}^{K_1} \sum_{k_2 = 1}^{K_2} b_{pk_1, pk_2} B_{k_1, 1}(t) B_{k_2, 2}(u)$, where $B_{k_1, 1}(t)$ and $B_{k_2, 2}(u)$ are pre-specified univariate spline basis functions and $b_{pk_1, pk_2}$ are spline coefficients.
Similar to FoSR, the penalized model reduces to a complicated mixed effects model and can be fit using existing software.
In practice, \verb|gam|-based software, including \verb|refund::pffr()|, does not allow missing data in the predictor $W_i(\cdot)$.
The solution we use is to smooth $W_i(\cdot)$ for each participant $i$ using FPCA and impute values that are missing on a specific grid (here $t = 1, 2, ..., 24$ weeks).

\subsection{Inference for Functional Coefficients}
\label{subsec:methods_ci}

We use a nonparametric bootstrap to estimate standard errors of functional coefficients, reducing reliance on distributional assumptions.
To obtain the standard error of $\widehat \beta_m(t)$ from the FoSR model in Equation \eqref{eq:fosr_eq}, for $b = 1, ..., B$, we (1) draw a sample of individuals with replacement, (2) run FPCA on $W_i(t)$ in this sample to estimate $\phi_k(t)$ used for expanding subject-specific effects $Z_i(t)$, (3) fit an FoSR model using the sample, and (4) extract the coefficient of interest at time $t$ as $\widehat \beta_m^{b}(t)$.
The standard error $\text{SE}\set{\widehat \beta_m(t)}$ is then estimated as the standard deviation of $\set{\widehat \beta_m^{b}(t) : b = 1, ..., B}$.
$B$ is typically large; we use $B = 300$.
To construct $\alpha$-level confidence intervals at each time point, one option is to use Wald intervals $\widehat \beta_m(t) \pm z_{(1-\alpha/2)} \times \text{SE}\set{\widehat \beta_m(t)}$.
However, this may inflate the type-I error rate due to the numerous time points. Moreover, the coefficient values at each point are inherently correlated due to the smoothness of the functional coefficient.
To remedy these issues, we use correlation and multiplicity adjusted (CMA) joint confidence intervals based on the nonparametric bootstrap of the max absolute statistic \citep{fda_with_r_crainiceanu, semiparametric_regression_ruppert2003}.
Specifically, for each bootstrap iteration, we calculate
$d_b = \max_{t \in \mathcal{T}} \frac{|\widehat \beta_m^{b}(t) - \frac{1}{B} \sum_b \widehat \beta_m^{b}(t)|}{\text{SE}\set{\widehat \beta_m(t)}}$.
Let $q_{(1 - \alpha)}$ denote the $(1 - \alpha)$th quantile of $d_b$ over $b = 1, ..., B$. The $1 - \alpha$ CMA confidence interval is constructed as
$\widehat \beta_m(t) \pm q_{(1 - \alpha)} \times \text{SE}\set{\widehat \beta_m(t)}$.
Furthermore, the pointwise $p$-value for $\widehat \beta_m(t)$ at time $t$ is the smallest value of $\alpha$ such that the $1 - \alpha$ CMA confidence interval at time $t$ does not include 0. Similarly, the global $p$-value for the entire coefficient function $\widehat \beta(\cdot)$ is the smallest value of $\alpha$ such that the $1 - \alpha$ CMA confidence interval does not reach 0 at any time (i.e. the minimum of the pointwise $p$-values).

For contrasts between coefficients, we adapt the same procedure, using $\widehat \beta_{m_1}(t) - \widehat \beta_{m_2}(t)$ in place of $\widehat \beta_{m}(t)$.
The same procedures are applied to construct CMA confidence intervals for the functional coefficients $\gamma_{m}(\cdot)$ and their contrasts from FPCA + regression.
Extending this procedure to FoFR is straightforward; the standard error of $\widehat \beta_p(t, u)$ is obtained by calculating the standard deviation of $\set{\widehat \beta_p^b(t, u): b = 1, ..., B}$.
Similar to FoSR, $d_b = \max_{t, u} \frac{|\widehat \beta_p^{b}(t, u) - \frac{1}{B} \sum_b \widehat \beta_p^{b}(t, u)|}{\text{SE}\set{\widehat \beta_p(t, u)}}$, and the CMA confidence interval is constructed accordingly.
Supplementary Section \ref{supp:interference} presents a modified bootstrap procedure that accounts for correlation induced by participants receiving the competitive and collaborative interventions in cohorts of three.

\section{Results}
\label{sec:results}

We applied the proposed functional regression framework to the STEP UP dataset to quantify PA intervention effects over the study periods.
Due to software constraints, participants with fewer than 4 observations in the period of interest were excluded in the analysis, resulting in $n = 580$ and $n = 367$ participants for the intervention and follow-up period, respectively.
In FPCA + regression, we first used all covariates as described in Section~\ref{subsec:covariates}.
In FoSR and FoFR, to increase computation speed, we included time-varying effects of study arm, and time-invariant effects of baseline steps, start day, age, and gender.
When comparing FoSR and FPCA + regression, we refit the FPCA + regression model with the same covariates as FoSR.
All analyses were conducted in R 4.4.0 \citep{Rsoftware}.

\subsection{Effectiveness of the PA Intervention}
\label{subsec:results_effectiveness}

To address this first question, we fit the models introduced in Sections~\ref{subsec:methods_fosr} and \ref{subsec:methods_fpca} for the intervention and follow-up periods separately. Here we focus on the results for the intervention period, and the results for the follow-up period are shown in Supplementary Section \ref{supp:follow_up}.
To further evaluate the models, simulations in Section \ref{sec:sim_study} assess the performance of FoSR and FPCA + regression.

\subsubsection{Function-on-scalar regression (FoSR)}
\label{subsubsec:results_fosr}

\begin{figure}
    \centering
    \includegraphics[width=\linewidth]{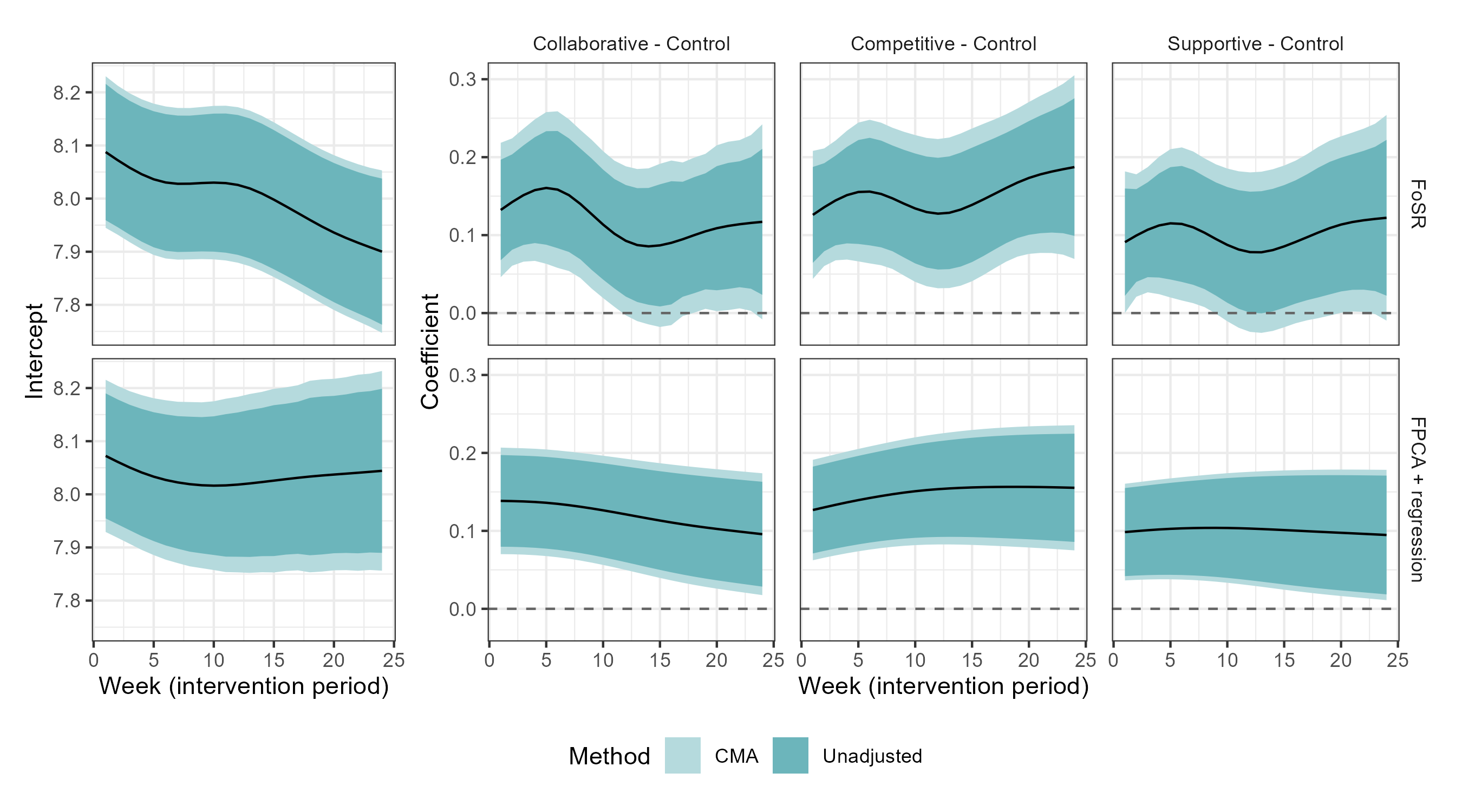}
    \caption{Functional coefficients from FoSR and FPCA + regression with unadjusted and correlation and multiplicity adjusted (CMA) confidence intervals. The coefficients show the effect of study arm relative to control. The effect at time $t$ is considered significant if the CMA CI does not contain 0. Note the outcome is log steps.}
    \label{fig:combined_int_coef_plot}
\end{figure}

Fitting FoSR using \verb|gam|-based approaches requires choosing the smoothing parameter $\lambda$.
We used fast REML (fREML) to select $\lambda$, as other options were too slow to fit the model on a standard laptop in under an hour.
The functional intercept $\beta_0(t)$ and coefficients $\beta_m(t)$ from equation \eqref{eq:fosr_funccoef} are plotted in the top row of Figure~\ref{fig:combined_int_coef_plot} along with 95\% unadjusted (pointwise) and CMA confidence intervals.
The functional intercept is the mean log steps by time for the control arm, adjusted for covariates; it is generally decreasing over the intervention period but remains constant around weeks 7 to 12.
The coefficient $\beta_m(t)$ is the effect of arm $m$ at time $t$ relative to control at time $t$, adjusting for covariates.
All three intervention arms have $\beta_m(t)$ above 0 for the full intervention period, indicating that they have higher expected step counts at all time points compared to control.
The effect is significant at the .05 level where CMA confidence intervals exclude 0: weeks 1--11 and 18--23 for the collaborative arm, all 24 weeks for the competitive arm, and weeks 1--8 and 20--22 for the supportive arm.
The effect of the collaborative and supportive arms at the other time points is weaker. In other words, the time periods during which the intervention is effective vary by arm. This is not surprising given how different these gamification strategies are. However, to our knowledge, this is the first time these results have been reported in STEP UP, highlighting how functional modeling can reveal additional temporal insights not captured by conventional methods. Additional sensitivity analyses (Supplementary Section \ref{supp:sensitivity}) suggest that the collaborative arm effects around weeks 18--23 and the supportive arm effects around weeks 20--22, while marginally significant, vary with preprocessing decisions and covariates. Nevertheless, our general conclusions that the effect is significant for the first portion of the intervention period in the collaborative and supportive arms and for the full period in the competitive arm remain robust.

The contrasts $\beta_{m_1}(t) - \beta_{m_2}(t)$ of coefficients and their confidence intervals are shown in Supplementary Figure $\ref{fig:s_combined_contrast_plot}$; no effects are significant at any time.
However, the competitive arm coefficient is higher than supportive for the entire intervention period and higher than collaborative for week 7 onward.
Furthermore, it appears that the efficacy of the collaborative intervention decreases over the intervention period relative to the other two arms.

As for the covariates, start day is significantly negatively associated with steps ($\beta = -.0009, p = .0015$).
This may be because the most eager participants signed up first, or it may be a seasonal effect as people are generally less active in colder months \citep{tucker2007_seasonPAreview}.
Baseline steps ($\beta = 9.20 \times 10^{-5}, p < .0001$) and age ($\beta = .0024, p = .0273$) are significantly positively associated with steps, and gender is not associated with steps ($\beta = .0354, p = .144$).
Note that STEP UP has a restricted age range and only moderate sample size.
The results for the follow-up period as well as the pointwise CMA $p$-values for both periods are provided in the Supplementary Materials (Figures \ref{fig:s_fosr_pvalue_plot}, \ref{fig:s_combined_coef_plot_post}, \ref{fig:s_all_contrast_plot_post}).

\subsubsection{FPCA + regression}
\label{subsubsec:results_fpca}

\begin{figure}
    \centering
    \includegraphics[width=\linewidth]{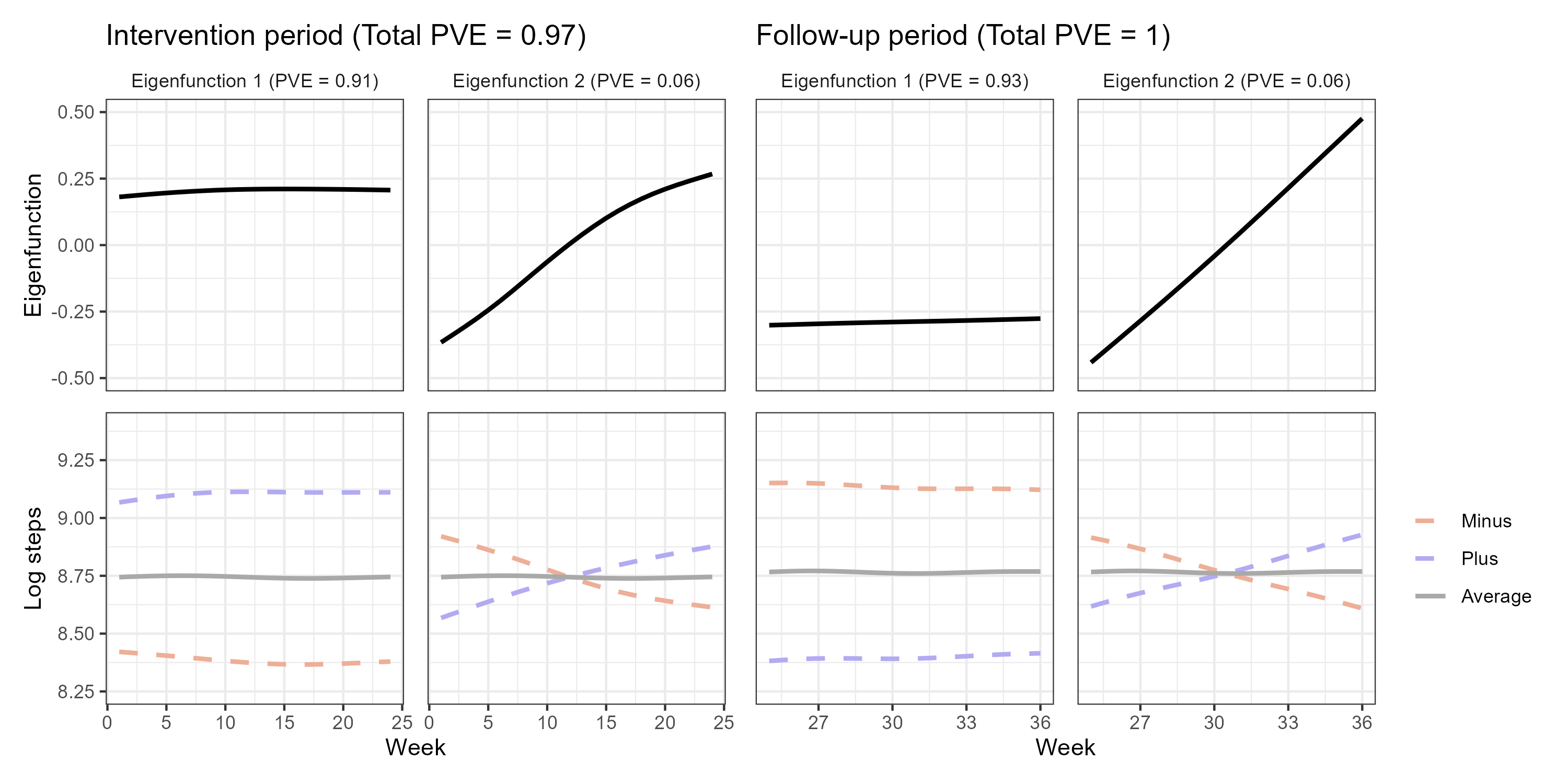}
    \caption{Eigenfunctions from FPCA. For each eigenfunction $\phi_k(\cdot)$, the top panel shows the eigenfunction with its percent variability explained (PVE), and the bottom panel shows the smoothed population average plus/minus a multiple of the eigenfunction, $\mu(t) \pm \sqrt{\lambda_k} \phi_k(t)$. For both periods, the first eigenfunction can be interpreted as a vertical shift, i.e., representing overall magnitude of steps, and the second eigenfunction roughly represents whether the trajectory is increasing or decreasing.}
    \label{fig:fpca_int_post}
\end{figure}

The eigenfunctions $\phi_k(t)$ from FPCA are shown in Figure~$\ref{fig:fpca_int_post}$; for both periods, $K = 2$ eigenfunctions are needed to achieve 95\% PVE.
For the intervention period, the first eigenfunction is positive and roughly constant over time.
Those with high scores on the first eigenfunction tend to have a higher overall number of steps compared to those with low scores.
The second eigenfunction is increasing from negative to positive along the domain, indicating that those with high scores increase steps over time, whereas those with low scores tend to decrease steps over time.
Since eigenfunctions are identifiable only up to multiplication by -1, the follow-up period eigenfunctions have roughly the same interpretation as that for the intervention period.

For the intervention period, the associations between scores and predictors after correcting for multiple testing are shown in the Supplementary Table \ref{table:inteffull}.
Baseline step count ($\beta = 4.09\times 10^{-4}, p < .001$), each of the three intervention arms (collaborative arm: $\beta = .54, p = .004$; competitive arm: $\beta = .71, p < .001$; supportive arm: $\beta = .49, p = .018$), and ESE ($\beta = .30, p = .004$) are significantly positively associated with scores on the first eigenfunction, indicating that these factors are associated with high overall step counts.
Start day is significantly negatively associated with scores on the first eigenfunction ($\beta = -.0051, p = .004$); starting later is associated with lower overall step counts, similar to FoSR.
Baseline step count is also significantly negatively associated with scores on the second eigenfunction ($\beta = -3.1 \times 10^{-5}, p < .001$), so those with high baseline steps tend to have decreasing step counts.

\subsubsection{Comparison between FoSR and FPCA + regression}
\label{subsubsec:results_comparison}

\begin{figure}
    \centering
    \includegraphics[width=.65\linewidth]{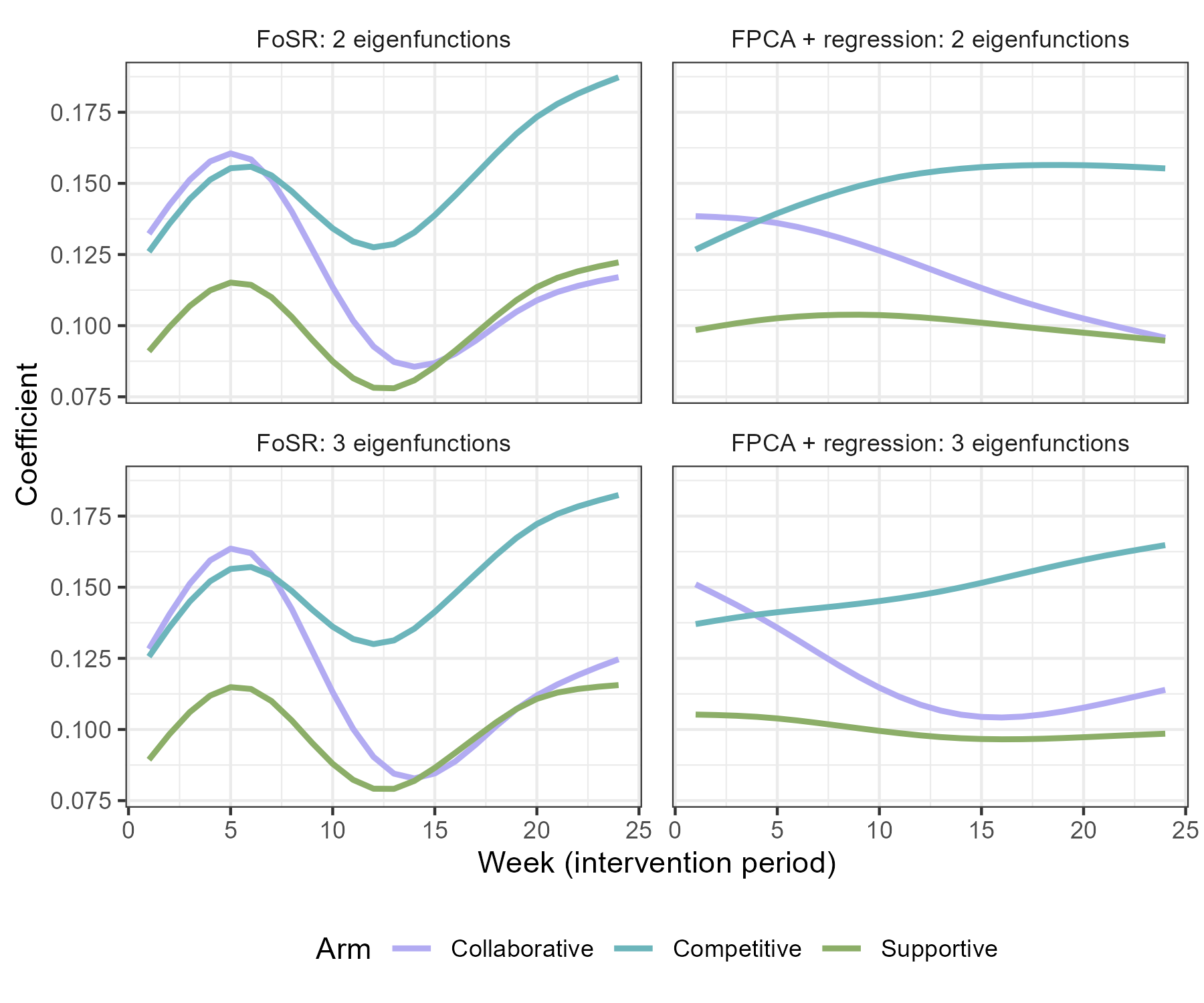}
    \caption{Functional coefficients from FoSR and FPCA + regression with $K = 2$ and $K = 3$ eigenfunctions. For FoSR, the eigenfunctions are the basis of random effects, and for FPCA + regression, regressions are done on scores from the first $K$ eigenfunctions.
    The shape of the coefficients, namely their concavity, changes by the number of eigenfunctions for FPCA + regression (right column), while the coefficients stay relatively similar for FoSR (left column).}
    \label{fig:combined_coef_plot}
\end{figure}

The functional coefficients $\gamma_m(t)$ and intercept $\gamma_0(t)$ from the two-step FPCA + regression model $\eqref{eq:fpca_reg_eq}$ are plotted on the bottom row of Figure~\ref{fig:combined_int_coef_plot}.
Based on the CMA confidence intervals, all three intervention arms have
significantly higher expected steps at all time points compared to control.
The functional coefficients from the two methods, overlaid in Figure \ref{fig:combined_coef_plot} for ease of comparison, share trends in magnitude.
In particular, the coefficients for all three intervention arms are greater than 0 for the entire intervention period.
Also, the coefficients for competitive and collaborative start out higher than the supportive, but the coefficient for collaborative arm drops to that of the supportive arm eventually.

However, the coefficient function shapes are visibly different between the two approaches, as the two-step model exhibits overly smooth patterns. This limitation arises because the coefficients in FPCA + regression are constrained to the surface spanned by the first two eigenfunctions, both of which are very smooth as shown in Figure~\ref{fig:fpca_int_post}. As a result, the two-step model lacks the flexibility to capture the more complex, time-varying effects with higher curvature.
In contrast, the FoSR approach allows greater flexibility by using more basis functions; for example, here 20 are used for estimating the coefficients $\beta_m(t)$.

In addition, the two-step model coefficients are more sensitive to the choice of $K$.
Figure~\ref{fig:combined_coef_plot} shows the functional coefficients using $K = 2$ (top row) or $K = 3$ (bottom row) eigenfunctions for both FoSR (left column) and two-step (right column) models.
For FPCA + regression, the shapes of the functions are noticeably different across panels, and we observe a change of concavity from $K = 2$ to $K = 3$.
In contrast, the shape of FoSR coefficients remains unchanged when using 3 eigenfunctions for the random effects.
This is expected because the eigenfunctions are used only to expand the random intercept $Z_i(t)$.

In summary, our results suggest all three intervention strategies have time-varying effects on PA, with the strongest effect observed near the beginning of the intervention.
Furthermore, the proposed FoSR model identifies additional time-varying patterns that cannot be captured by the two-step model. While the FoSR model can be computationally intensive, software advancements facilitate its implementation on moderately sized datasets such as STEP UP.

\subsection{Sustainability of PA Intervention}
\label{subsec:results_fofr}

\begin{figure}
    \centering
    \includegraphics[width=\linewidth]{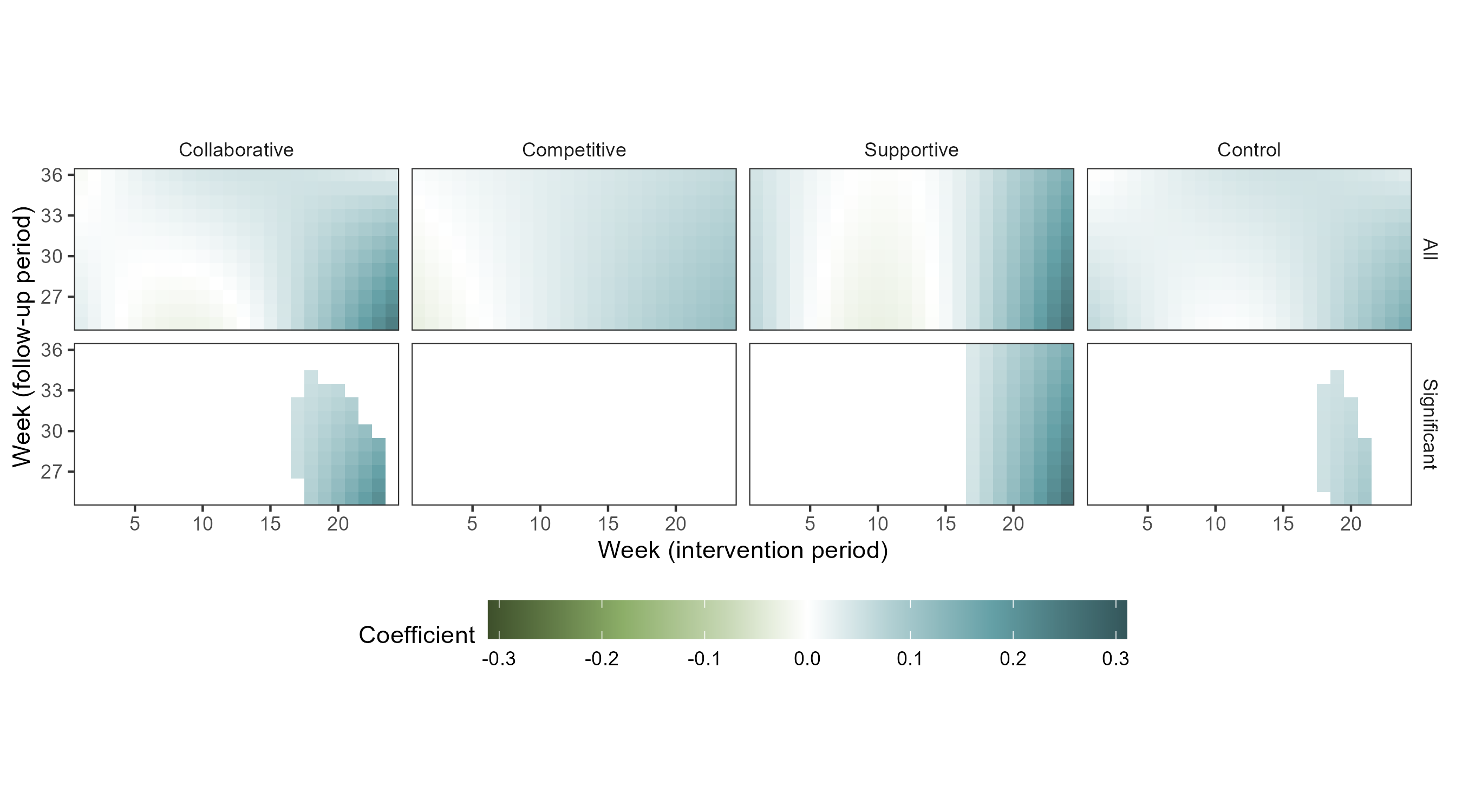}
    \caption{Functional coefficients $\beta_p(t, u)$ from the FoFR model. The shade of the tile at week $t$ in the intervention period and week $u$ in the follow-up period corresponds to the effect of steps at week $t$ on steps at week $u$ for that study arm, adjusted for baseline steps.
    In the second row of plots, the tile at $(t, u)$ is masked white if the effect is not significant, that is, the CMA CI for $\beta_p(t, u)$ includes zero.}
    \label{fig:fofr_coef_x}
\end{figure}

Next, we fit the FoFR model in Section~\ref{subsec:methods_fofr} to evaluate the association between activity in the intervention and follow-up periods. The coefficients $\beta_p(t, u)$ for the four arms are shown in Figure~\ref{fig:fofr_coef_x}.
For the collaborative arm, the end of the intervention period is significantly associated with the start of the follow-up period (dark blue in the bottom right). That is, those with high step counts at the end of the intervention are more likely to have high step counts going into the follow-up, but the effect diminishes over the follow-up period.
In contrast, for the supportive arm, the end of the intervention period is significantly associated with the entire follow-up period, indicating that those with high PA at the end of the intervention sustained it throughout the follow-up.
For the control arm, the association between intervention and follow-up period steps is more uniform over both domains, which is reasonable as there was no difference between intervention and follow-up for control.
The competitive arm coefficient exhibits less variation over both domains than the other arms.

\begin{figure}
    \centering
    \includegraphics[width=.9\linewidth]{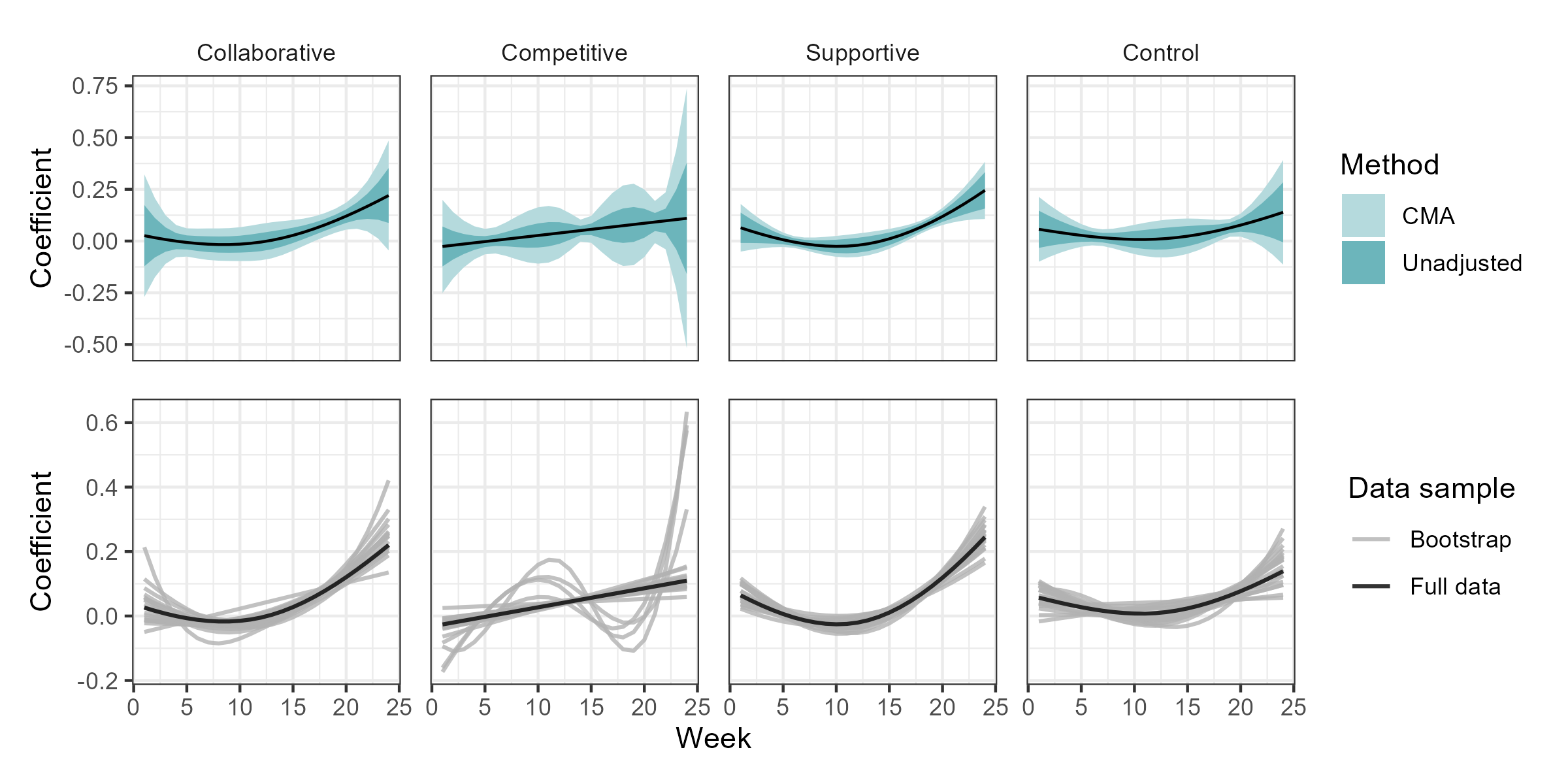}
    \caption{Cross section of FoFR coefficients $\beta_p(t, u)$ at $u = 26$. The top row includes CMA confidence intervals and the bottom row includes the estimates from 20 bootstrap samples. Some coefficients in some samples are estimated to be linear, resulting in the cinching pattern for the confidence intervals.}
    \label{fig:combined_bs_cross_sec_plot_26}
\end{figure}

For further insights, Figure~\ref{fig:combined_bs_cross_sec_plot_26} plots a cross-sectional view, fixing follow-up time at $u = 26$.
Mostly, $\beta_p(t, u)$ is increasing over $t$, indicating that the association between intervention steps and steps at week $u = 26$ is stronger toward the end of the intervention.
The collaborative arm confidence interval is wider at both tails of the intervention; this is why the bottom right corner in Figure~\ref{fig:fofr_coef_x} is not significant.
The cinching pattern in the confidence intervals for the collaborative, control, and especially competitive arms is due to the linearity of the estimated coefficients in several bootstrap iterations.
This indicates the smoothing parameter estimated by fREML was sometimes very large, which warrants further investigation but is beyond the scope of this paper.
In the bootstrap iterations where the competitive arm coefficient is nonlinear, there is a negative association between steps at the start of the intervention and the start of the follow-up, perhaps because eager participants start enthusiastically but struggle to maintain step counts after the intervention.

As for covariates, start day is significantly negatively associated with follow-up period steps ($\beta = -.0009, p < .001$), and so is age ($\beta = -.0020, p = .031$). Similar to FoSR, gender is not significantly associated with follow-up period steps ($\beta = -.0128, p = .532$). Unlike FoSR, baseline steps in the FoFR model also show no significant association ($\beta = .0146, p = .271$), possibly due to correlation with intervention period steps, a predictor in FoFR.

\section{Simulation Study}
\label{sec:sim_study}

\subsection{Methods}

We conduct a simulation study to assess the performance of the FoSR model and compare it with the traditional FPCA + regression method.
The data are generated using the following steps:
\begin{enumerate}
    \item Conduct FPCA on the full intervention period data and extract the top 2 eigenfunctions $\widehat \phi_k(t)$ and corresponding eigenvalues $\widehat \lambda_k$.
    \item Fit FoSR model \eqref{eq:fosr_funccoef} to the full intervention period data and extract the fitted intercept $\widehat \beta_0(t)$, the fitted coefficients $\widehat{\beta}_m(t)$ for $m$ corresponding to the three intervention arms, the fitted time-invariant coefficients $\widehat b_m$ for $m$ corresponding to age, gender, baseline steps, and start day, and the variance of the error $\widehat \sigma_{\epsilon}^2$.
    \item Sample $N$ individuals with replacement from the data and extract their covariates $X_{im}$, specifically baseline steps, study arm, age, gender, and start day.
    \item Construct $\widetilde{W}_i(t)$ for this simulation sample according to
    $\widetilde{W}_i(t) = \widehat \beta_0(t) + \sum_{m = 1}^Q X_{im} \widehat \beta_m(t) + \sum_{m = Q + 1}^M X_{im} \widehat{b}_m + \sum_{k= 1}^K \zeta_{ik} \widehat \phi_k(t) + \epsilon_i(t),$
    where $\zeta_{ik}$ are independent $ N(0, \widehat \lambda_k)$ and $\epsilon_i(t)$ are independent $N(0, \widehat \sigma_{\epsilon}^2)$.
    \item For each $\widetilde{W}_i(t)$, sample an individual $j$ from the real data and copy their missingness pattern by setting $\widetilde{W}_i(t)$ at time $t$ as missing if $W_j(t)$ is missing at that time.
\end{enumerate}

\begin{figure}
    \centering
    \includegraphics[width=0.65\linewidth]{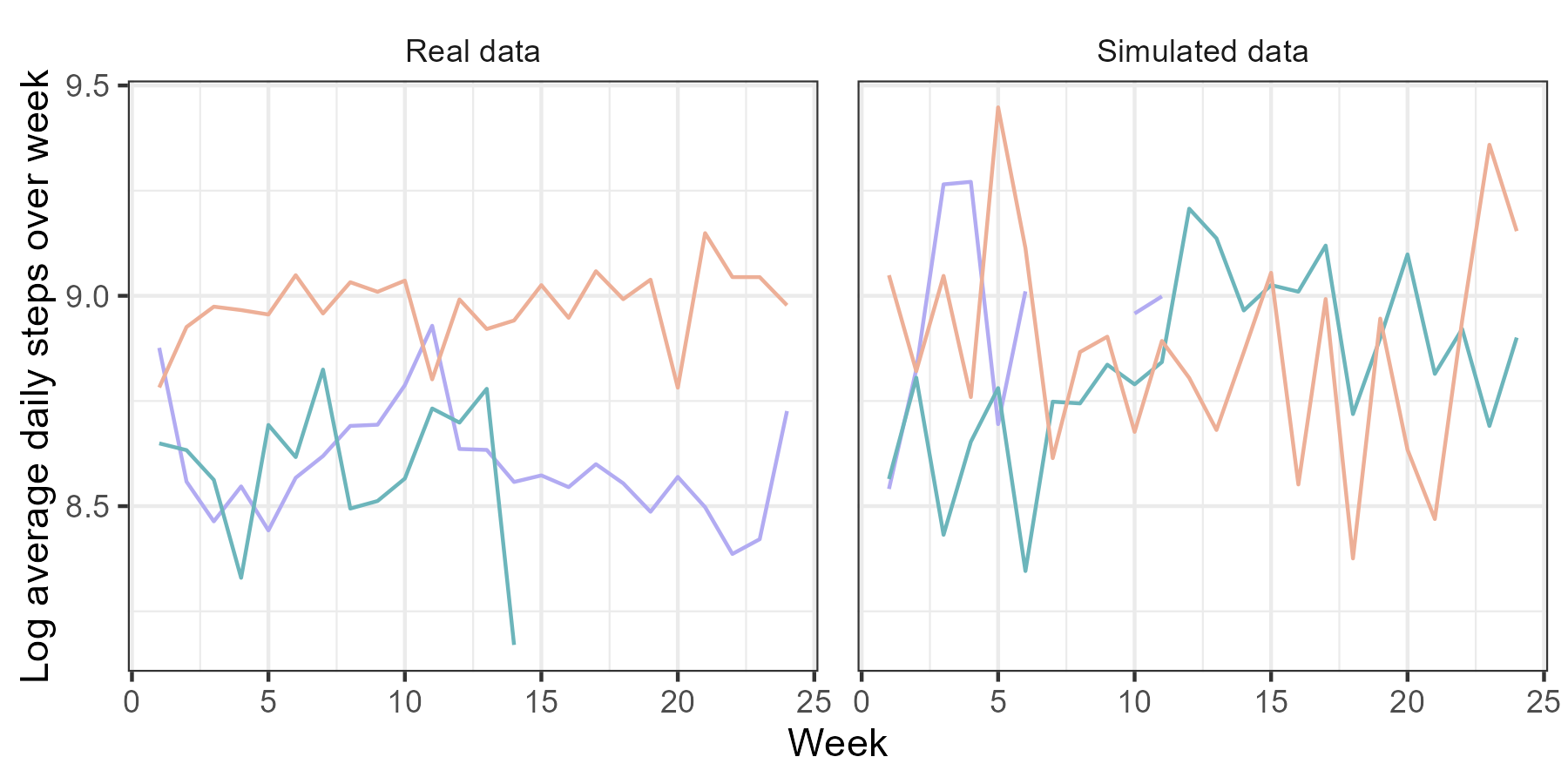}
    \caption{Intervention period step trajectory for three randomly sampled individuals in the real data and one simulated dataset. Both datasets show high within-subject variability and complex missingness patterns.}
    \label{fig:sim_truth}
\end{figure}

Figure \ref{fig:sim_truth} shows the PA trajectories of three randomly selected subjects from both the real data and simulated data. Both panels exhibit high within-subject variability and complex missingness patterns, indicating good performance of our data generation in mimicking real data.
For each simulated dataset, we fit the FoSR model (\ref{eq:fosr_funccoef}) and the FPCA + regression model (\ref{eq:fpca_reg_eq}). Then, we extract estimates for $\beta_m(t)$ and $\gamma_m(t)$ for the FoSR and FPCA + regression model, respectively.
2000 simulated datasets are generated and analyzed.

\subsection{Results}

\begin{figure}
    \centering
    \includegraphics[width=0.65\linewidth]{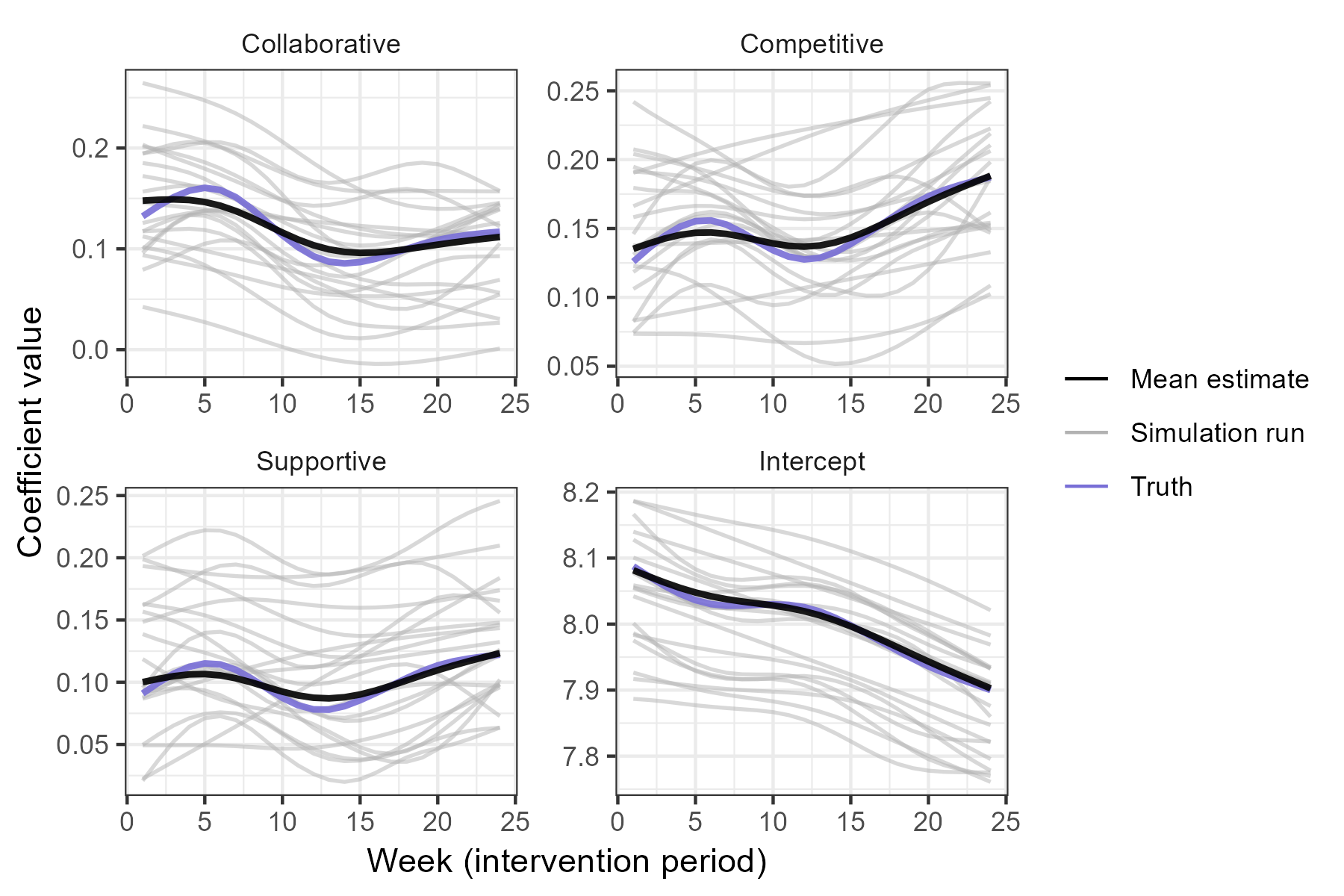}
    \caption{Estimated coefficients from FoSR simulation study. The mean estimate in black shows the average over 2000 simulation runs; estimated coefficients from 20 randomly selected simulations are plotted in gray.}
    \label{fig:sim}
\end{figure}

\begin{figure}
    \centering
    \includegraphics[width=.75\linewidth]{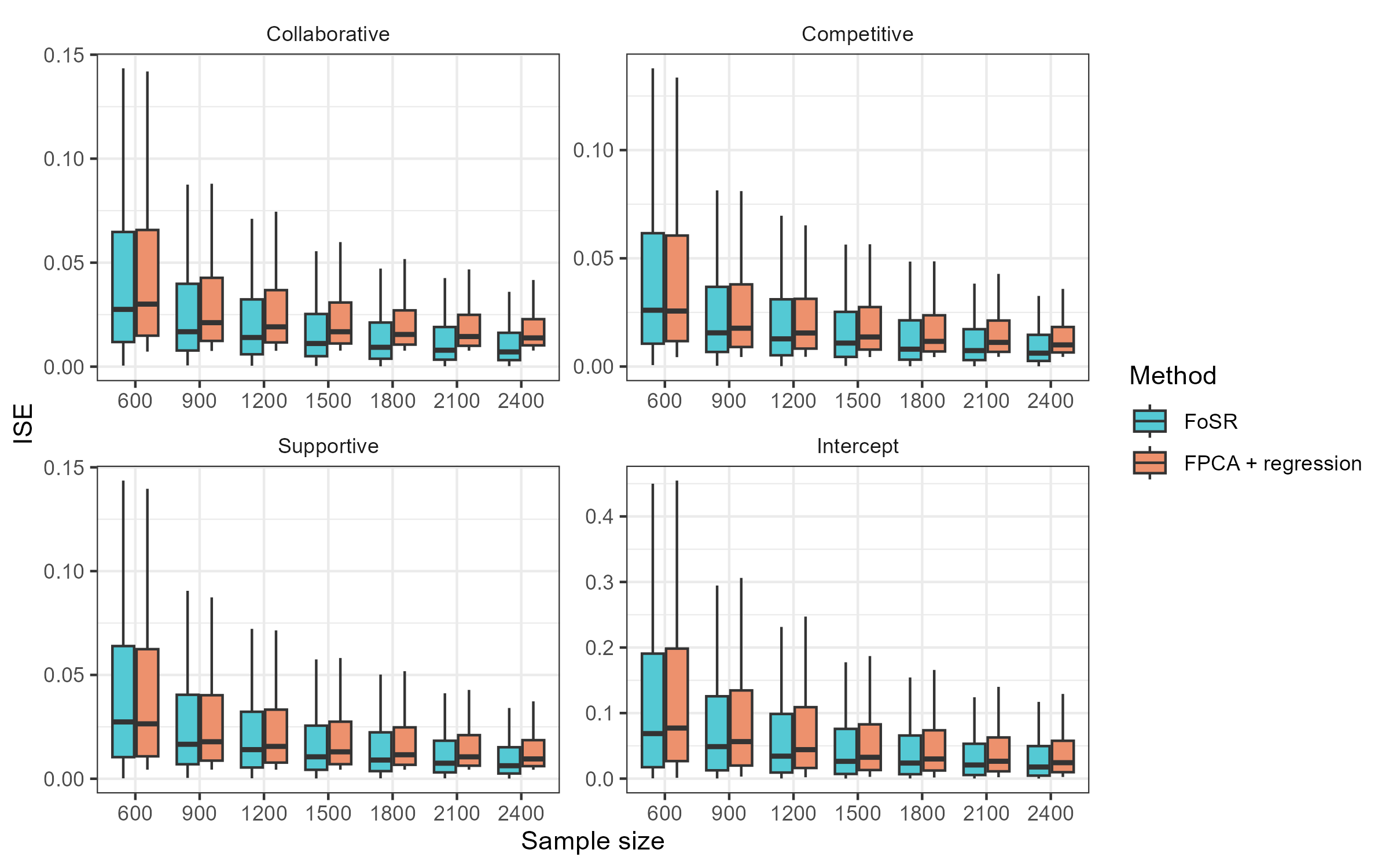}
    \caption{Integrated squared error (ISE) of coefficients in simulation by sample size and method. For coefficients corresponding to the intercept, intervention arm, and baseline steps, estimation accuracy improves with increasing sample size. Additionally, median ISE is lower using FoSR compared with FPCA + regression, especially as sample size increases.
    Outliers are omitted.}
    \label{fig:sim_ise}
\end{figure}

For the intercept and $\beta_m(t)$ corresponding to the three intervention arms, Figure \ref{fig:sim} compares the true coefficient (purple) to the mean estimated coefficient over 2000 simulations (black).
Overall, the mean estimated coefficient is relatively close to the true coefficient, and the shapes of the estimated coefficients are roughly similar to the truth.
Figure \ref{fig:sim_ise} displays the distribution of integrated squared error (ISE) of the estimated coefficients by sample size and method. In general, the estimation accuracy improves as sample size increases. Additionally, the median ISE from FoSR is consistently lower than FPCA + regression, especially when the sample size is large.
These results suggest that the proposed framework is capable of capturing non-linear effects across variables, is robust to changes in model parameters, and outperforms traditional methods.

\section{Discussion}
\label{sec:discussion}

We introduced a functional regression framework to quantify the time-varying intervention effects in PA studies with high-dimensional endpoints. Notably, we showed that the common two-step ``FPCA + regression" approach, although widely used in applications, imposes strong constraints to the coefficients and is
oversensitive to eigenfunctions.
To address these limitations, we proposed a function-on-scalar regression (FoSR) model and extended it to a function-on-function regression (FoFR) model, both of which offer more flexible estimations and require minimal effort to implement. We also implemented a simple and valid inferential approach, and simulations demonstrated the promising performance of our methods.

Applying our FDA approach to STEP UP uncovered strategy-specific temporal patterns overlooked in prior analyses that used scalar summary measures.
In particular, the intervention effect was stronger for the collaborative and supportive arms towards the beginning of the intervention and waned towards the end. In contrast, the effect of the competitive arm was sustained over the entire intervention period.
To our knowledge, this is the first time such inferential results have been revealed and reported in the STEP UP study.
In addition, we modeled the associations of PA between the intervention period and the follow-up period, a question not considered previously, highlighting FDA's potential to reveal cross-period dynamics that can inform the design of interventions aimed at promoting long-term PA.

Several extensions could yield further insights.
First, we adjusted for calendar time by adding participant start day as a covariate. However, this does not capture the cyclical effects of season. Because all participants started between February and July 2018, disentangling the effects of season vs. time in the study is difficult yet potentially informative.
Second, the FoFR model can be extended to produce subject-specific prediction intervals for the follow-up period, a step toward personalized interventions.
Third, the competitive and collaborative interventions occurred in groups of three participants. Dependence among individuals in the same group may be accounted for through multilevel FoSR or FoFR models, or by modeling interference effects directly in a causal framework.
Finally, although STEP UP randomizes only at baseline, the framework can be extended to handle adaptive randomization designs.

This study also points toward several methodological and computational research directions to improve ease of fitting and interpreting functional regression models. First, the current FoFR software does not allow missing data in the functional predictor. Second, in both FoSR and FoFR, the estimated smoothing parameter for certain coefficients was sometimes extremely large, resulting in linear coefficients.
Third, model fitting with bootstrap inference can be computationally intensive for very large datasets, encouraging the development of ready-to-use analytic approaches.
These directions will be pursued in future work.

\noindent \textbf{Acknowledgments:} The authors would like to thank STEP UP participants, the Penn Medicine Nudge Unit and Deloitte for running the trial, and the Minnesota Supercomputing Institute at the University of Minnesota.

\noindent \textbf{Code availability:} Code to reproduce the analysis is available at \url{https://github.com/nidhipai/stepup_fda}.

\bibliographystyle{plainnat}
\bibliography{refs}

\clearpage

\setcounter{section}{0}
\setcounter{subsection}{0}
\setcounter{equation}{0}
\setcounter{figure}{0}
\setcounter{table}{0}

\renewcommand{\thesection}{S\arabic{section}}
\renewcommand{\thesubsection}{S\arabic{section}.\arabic{subsection}}
\renewcommand{\theequation}{S\arabic{section}.\arabic{equation}}

\renewcommand{\thefigure}{S\arabic{figure}}
\renewcommand{\thetable}{S\arabic{table}}

\begin{center}
\huge Supplementary Materials
\end{center}

\section{Sensitivity to preprocessing and covariates}
\label{supp:sensitivity}

To investigate the sensitivity of the intervention effect conclusions in Section \ref{subsec:results_effectiveness} to the preprocessing in Subsection \ref{subsec:preprocessing} and variable selection, we repeated the main FoSR and FPCA + regression analyses with alterations.

\begin{figure}[!htb]
    \centering
    \includegraphics[width=0.75\linewidth]{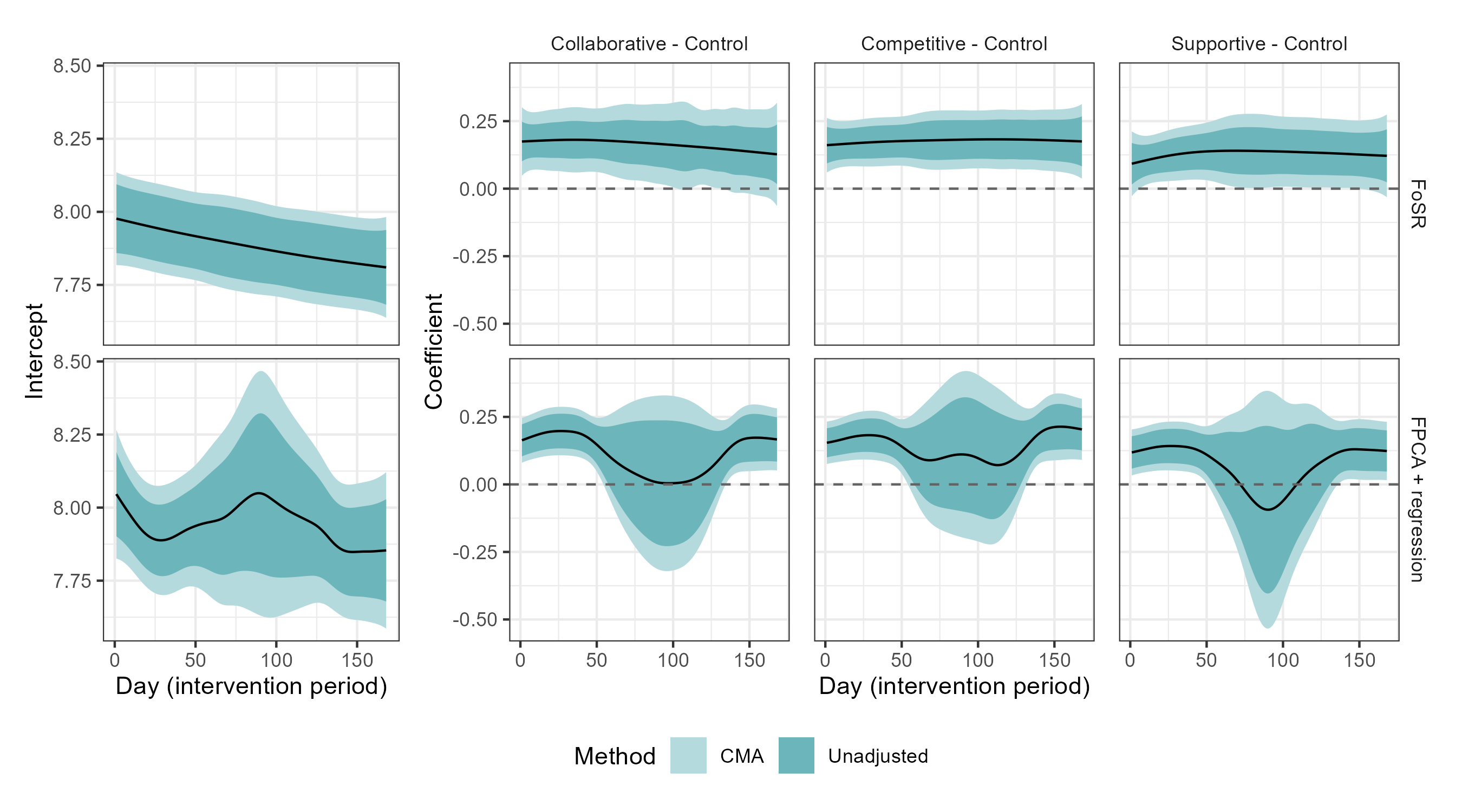}
    \caption{Sensitivity analysis of functional coefficients for the intervention period, using daily data instead of weekly average measures. Compared to their counterparts in Figure \ref{fig:combined_int_coef_plot}, the FoSR coefficients are flatter and the FPCA + regression coefficients have very different shapes.
    }
    \label{fig:s2daily_fig3}
\end{figure}

\begin{figure}[!htb]
    \centering
    \includegraphics[width=0.70\linewidth]{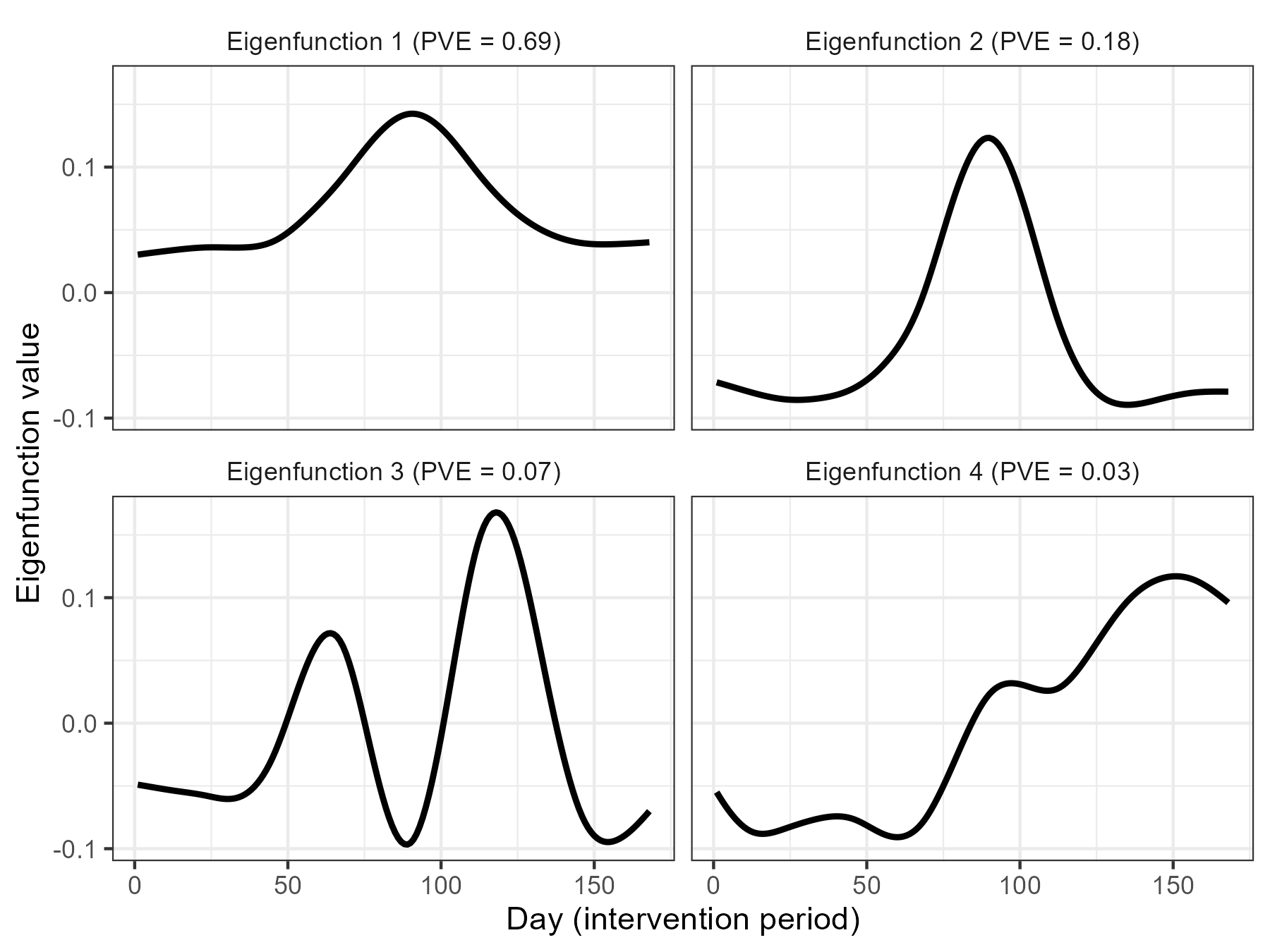}
    \caption{Eigenfunctions from FPCA with daily steps in the intervention period. The top four eigenfunctions are shown with their percent variability explained (PVE), together explaining 96.7\% of variability in the data. These differ substantially from the eigenfunctions using weekly averages (left side of Figure \ref{fig:fpca_int_post}) and result in the shapes of the FPCA + regression coefficients in Figure \ref{fig:s2daily_fig3}.}
    \label{fig:s_daily_4ef_plot}
\end{figure}

First, in the main analysis, daily steps were averaged into weekly steps. 
Participants took more steps on Saturdays compared to other days in the week, perhaps due to a different lifestyle schedule or the weekly structure of the game. This contributed to skewed residuals, violating the normality assumptions of the models used.
Since the primary interest is the longitudinal trajectory of steps over nine months, in the main analysis we remedied this issue by averaging steps over a week.
Alternatively, Figure~\ref{fig:s2daily_fig3} shows the results of fitting the FoSR and FPCA + regression model with daily instead of weekly data; it is analogous to Figure \ref{fig:combined_int_coef_plot}.
With daily data, four principal components are needed to explain 95\% of the variability, so the first $K = 4$ eigenfunctions form the basis for the functional random effects in FoSR and are the eigenfunctions used in FPCA + regression.
Compared to their counterparts in Figure \ref{fig:combined_int_coef_plot}, the FoSR coefficient (top row) are similar in magnitude but flatter.
In contrast, the FPCA + regression coefficients have a very different shape, mirroring the eigenfunctions (Figure \ref{fig:s_daily_4ef_plot}), which in turn are very different from the eigenfunctions in the weekly analysis (Figure \ref{fig:fpca_int_post}).
As discussed in Sections \ref{subsec:methods_fpca} and \ref{subsubsec:results_comparison}, FPCA + regression is very sensitive to the number and shape of eigenfunctions used, which may also be contributing to the wide confidence intervals, as coefficients vary significant between bootstrap samples.

\begin{figure}
    \centering
    \includegraphics[width=0.75\linewidth]{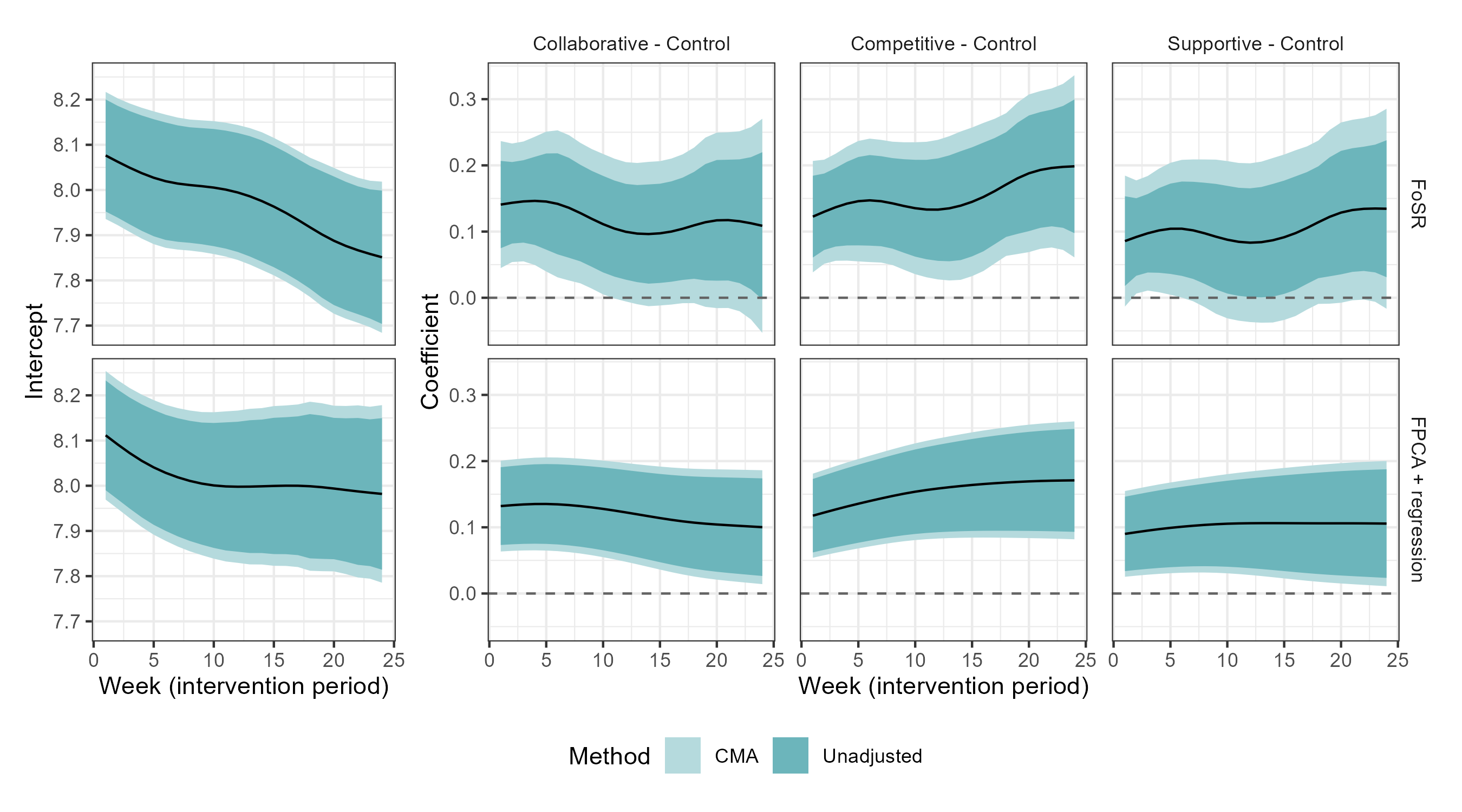}
    \caption{Sensitivity analysis of functional coefficients for intervention period, labeling a week's observation as NA only if there is no data for the week.
    Compared with Figure \ref{fig:combined_int_coef_plot}, the coefficients are slightly smoother, with wider confidence intervals.}
    \label{fig:s1NA_fig3}
\end{figure}

Second, in the main analysis, weeks with fewer than 3 days of data were labeled NA. To understand sensitivity, Figure \ref{fig:s1NA_fig3} shows the results when only weeks with no data are labeled NA.
The FoSR coefficients are smoother than those in Figure \ref{fig:combined_int_coef_plot}, and the confidence intervals are wider, particularly near the end of the intervention period.
This is possibly because weeks that are the average of only one or two days have more variability, and there are more such days toward the end of the intervention.

\begin{figure}[!htb]
    \centering
    \includegraphics[width=0.75\linewidth]{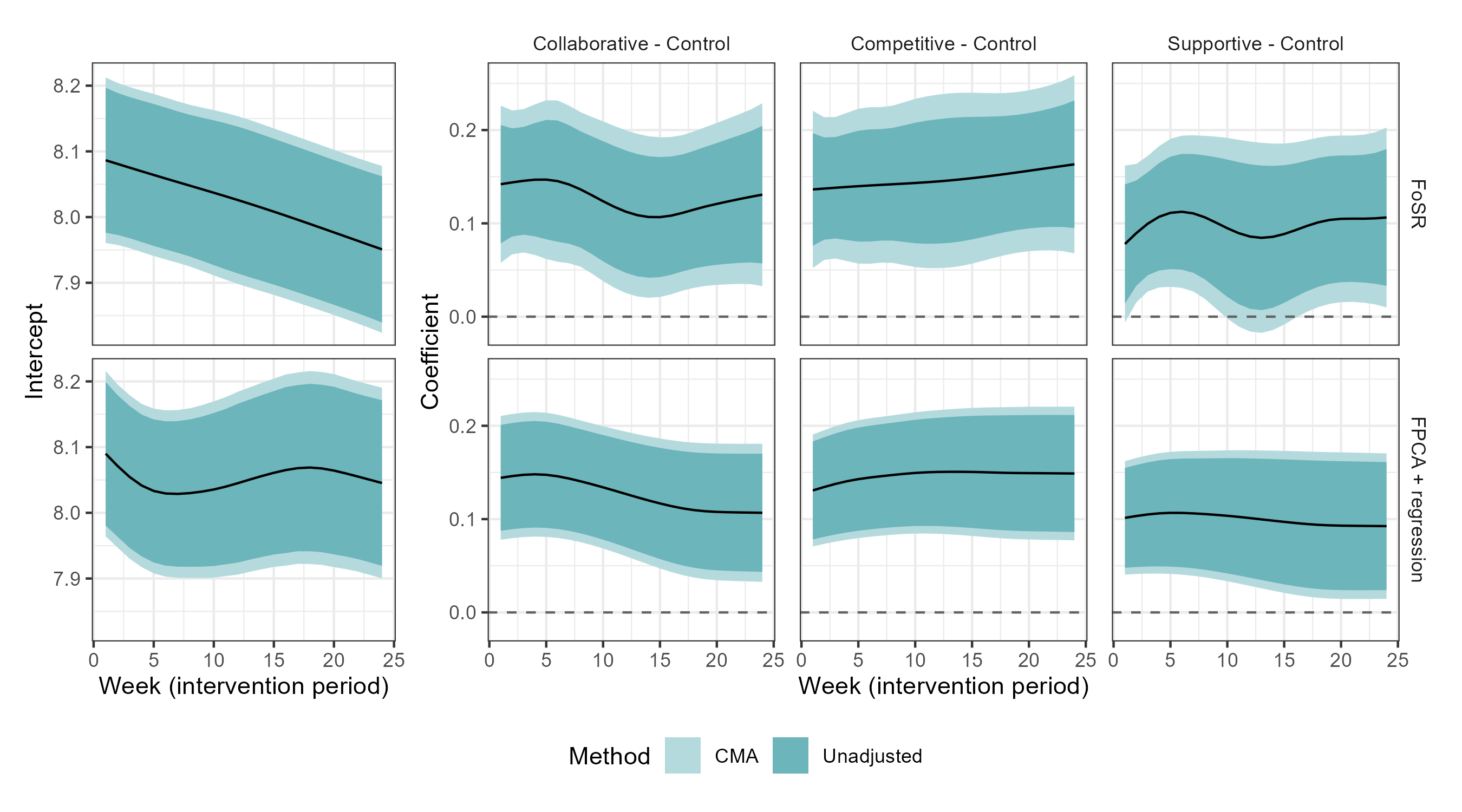}
    \caption{Sensitivity analysis of functional coefficients for intervention period, imputing data with FPCA before averaging by week. In the FoSR model, the intercept and coefficient functions for the collaborative and competitive arms are smoother (i.e. flatter) than their counterparts in Figure \ref{fig:combined_int_coef_plot}. The collaborative arm coefficient is above 0 the entire intervention period, a notable difference. The FPCA + regression functions are similar to those in Figure \ref{fig:combined_int_coef_plot}, except the intercept here is less smooth.
    }
    \label{fig:s15impute_fig3}
\end{figure}

As a second option, we imputed missing steps in the intervention period with FPCA via FACE \citep{face_xiao2016} before averaging daily steps (Figure \ref{fig:s15impute_fig3}). The FoSR intercept and competitive arm coefficient are much flatter than without the imputation. The FoSR collaborative arm coefficient retains its shape, but is significantly greater than 0 for the entire intervention period, whereas the confidence interval includes 0 at more points in the main analysis.
The FPCA + regression intercept is slightly less smooth, but the other FPCA + regression coefficients and confidence intervals are similar to the original results.
Note that FACE has some limitations for imputing data in this case: for weeks outside the range of the observed data for a subject, FACE begins by imputing the subject's average. Although it alternates between imputation and covariance estimation, there is both more missingness and lower step counts as the intervention period continues, possibly leading to systematically inflated imputed step counts.

\begin{figure}
    \centering
    \includegraphics[width=0.8\linewidth]{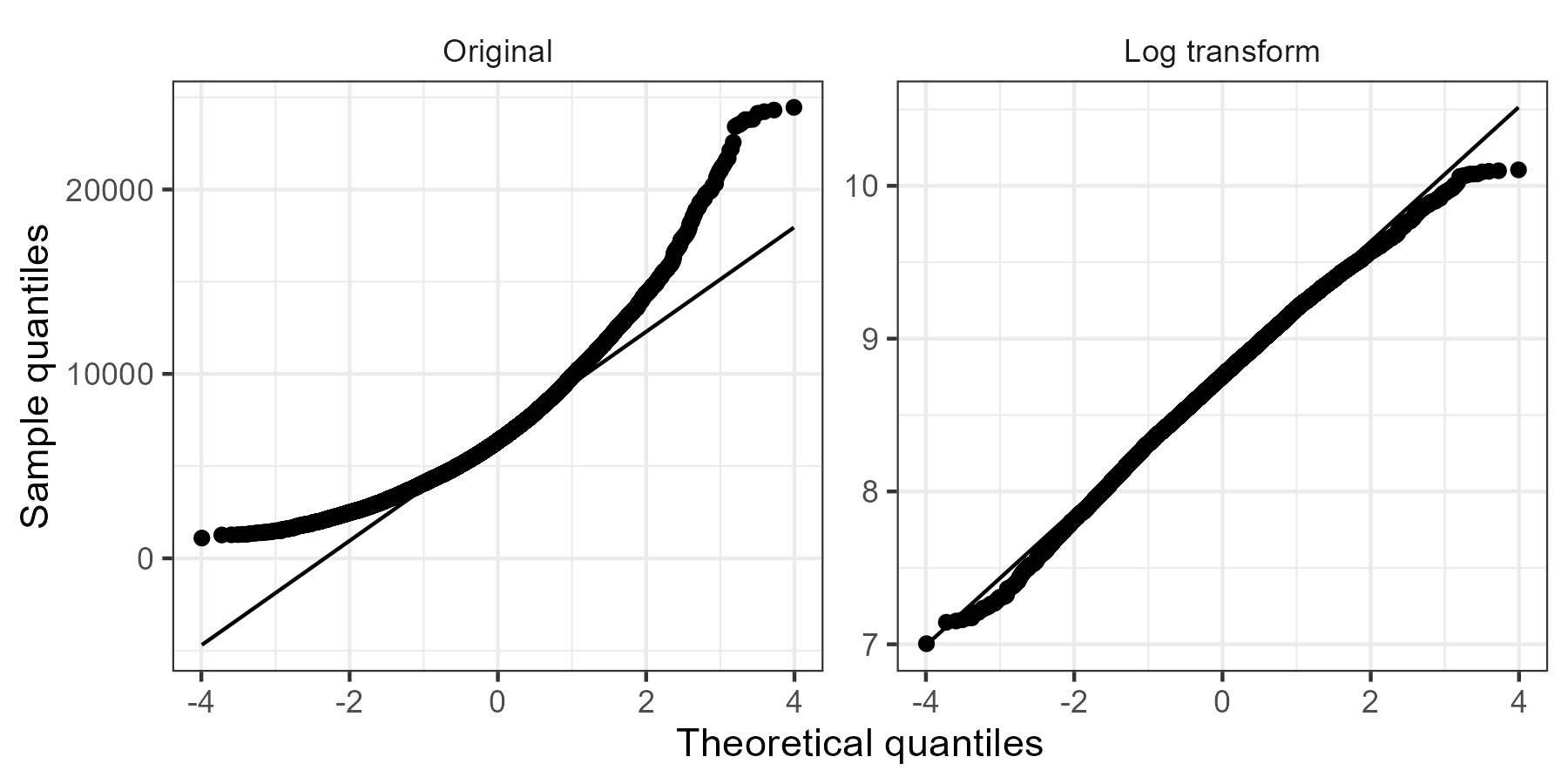}
    \caption{Normal quantile-quantile (Q-Q) plots of average weekly step counts with and without a log transform. The Q-Q plot on the left with the original data shows right-skewness of weekly average steps, and this is corrected via the log transform.}
    \label{fig:s_log_transform_qq}
\end{figure}

\begin{figure}[!htb]
    \centering
    \includegraphics[width=0.75\linewidth]{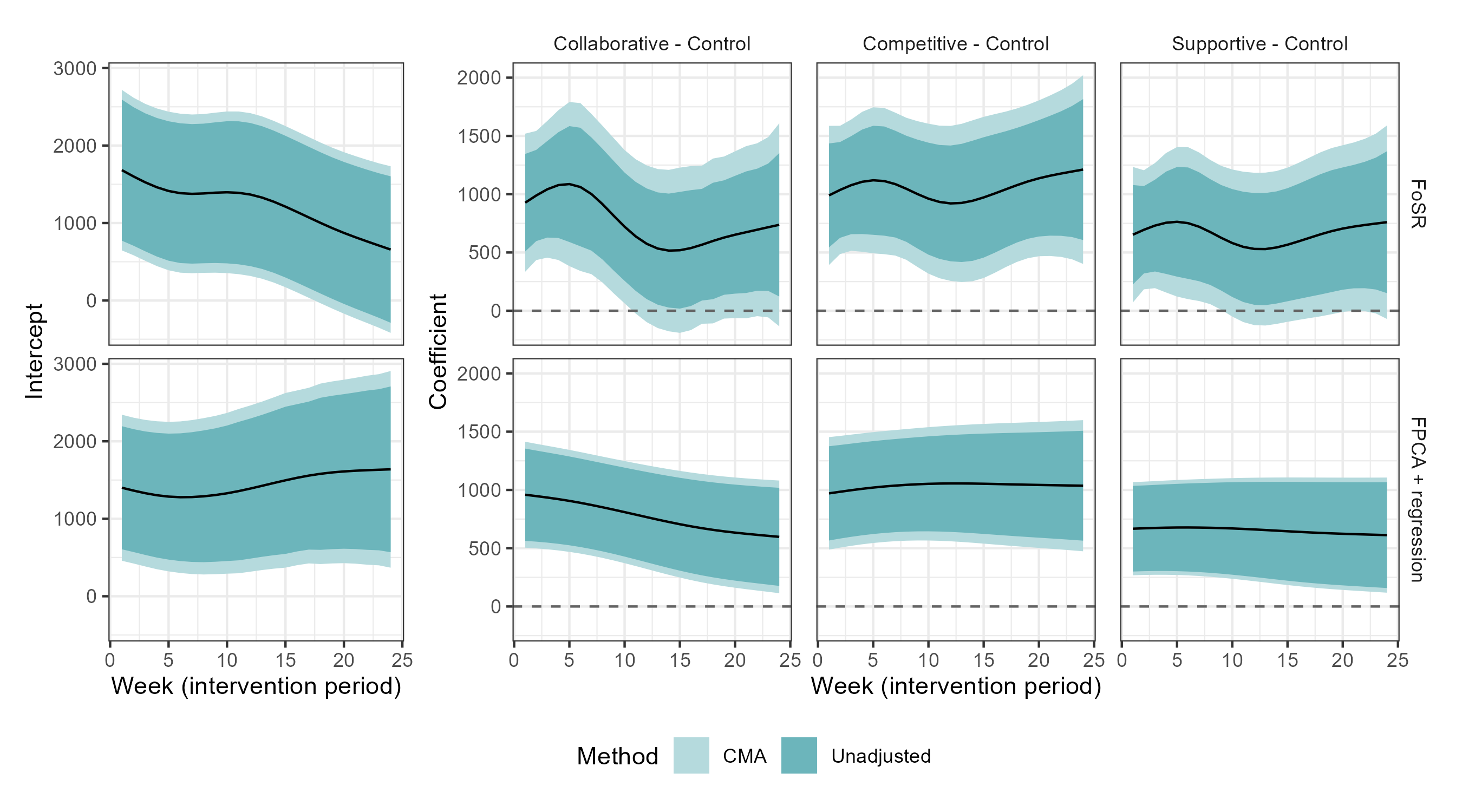}
    \caption{Sensitivity analysis of functional coefficients for intervention period, without a log transformation.
    Although the scale of the y-axis differs, the coefficient shapes are almost the same as Figure \ref{fig:combined_int_coef_plot}. Here, the confidence intervals are slightly wider, especially for the collaborative arm.}
    \label{fig:s3log_fig3}
\end{figure}

Third, the step counts and thus residuals were skewed, so we log transformed steps in the main analysis to satisfy normality assumptions (see Figure \ref{fig:s_log_transform_qq}).
However, Figure \ref{fig:s3log_fig3} shows that the results without the log transformation are very similar. The confidence intervals are slightly larger, so collaborative and supportive interventions effects are not significant at the end of the intervention period, but these effects were only borderline significant in the main analysis.
The lack of changes also indicates some robustness of the functional models to violations of normality.

\begin{figure}[!htb]
    \centering
    \includegraphics[width=0.75\linewidth]{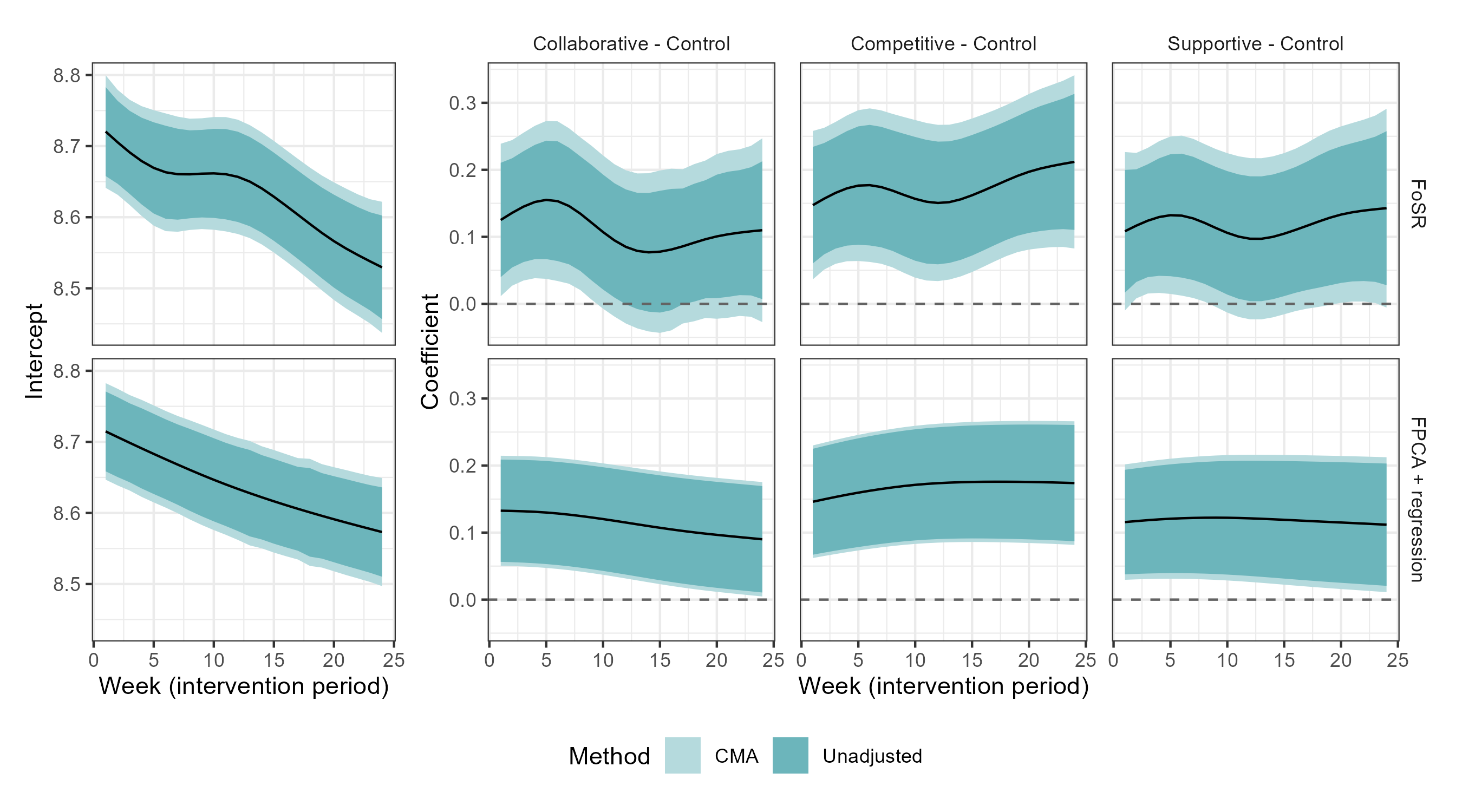}
    \caption{Sensitivity analysis of functional coefficients for intervention period, unadjusted for covariates. In both the FoSR and FPCA + regression models, the intercept is shifted up compared to Figure \ref{fig:combined_int_coef_plot}. The coefficient shapes and confidence intervals widths are otherwise similar.}
    \label{fig:s4unadj_fig3}
\end{figure}

\begin{figure}[!htb]
    \centering
    \includegraphics[width=0.75\linewidth]{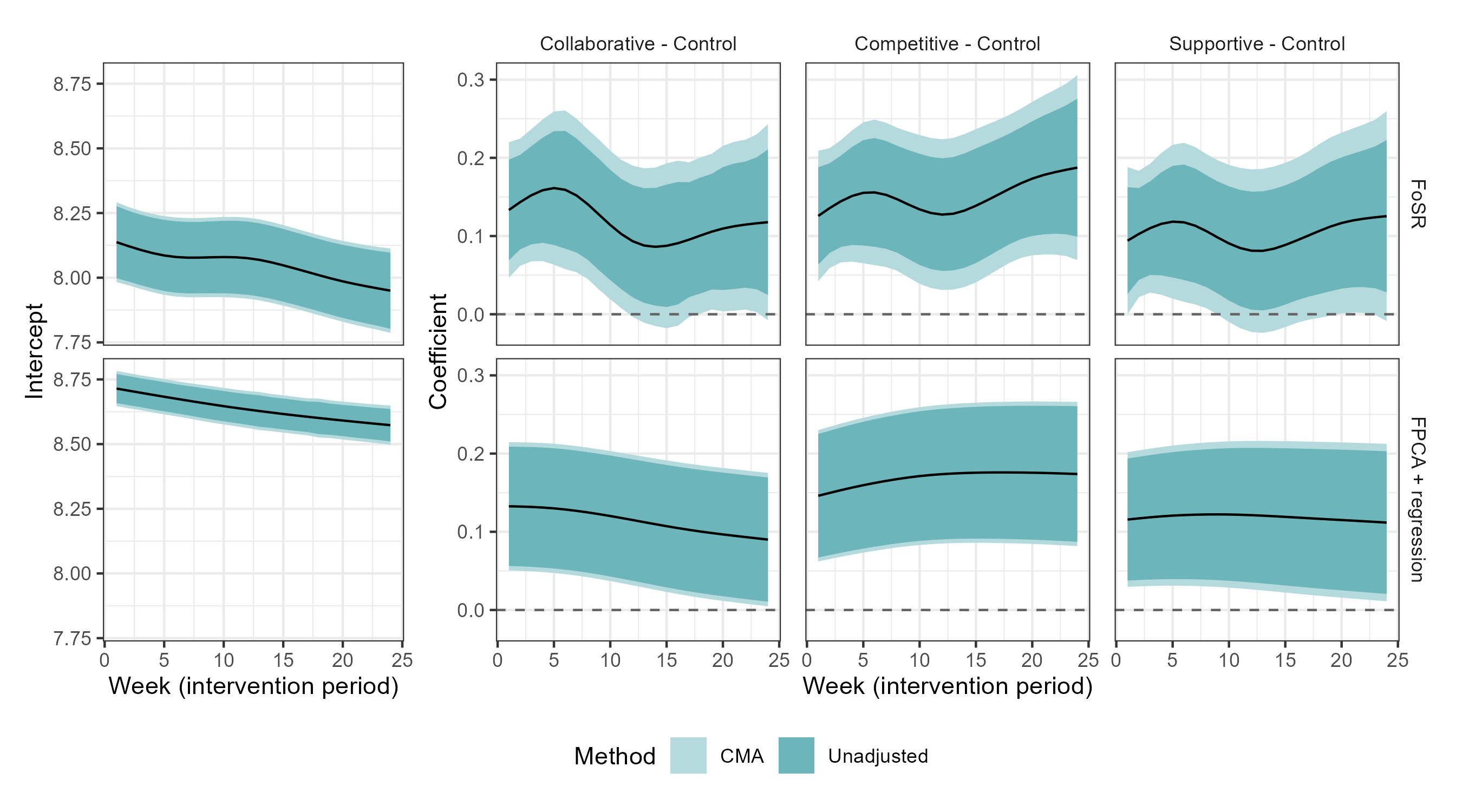}
    \caption{Sensitivity analysis of functional coefficients for intervention period, adjusted for additional covariates: race, marital status, and income. The intercept in the FPCA + regression model is larger than in Figure \ref{fig:combined_int_coef_plot} throughout the functional domain, possibly due to the reference levels of covariates; however, the intervention effect coefficients remain similar.
    }
    \label{fig:s4fullyadj_fig3}
\end{figure}

Finally, we checked the sensitivity of the conclusions to the variables included in the model. In the main analysis, we adjusted for baseline steps, start day, age, and gender with time-invariant effects. Figure \ref{fig:s4unadj_fig3} shows the results adjusting for no covariates, and Figure \ref{fig:s4fullyadj_fig3} shows the results adjusting also for race, marital status, and gender, so the covariates mirror those in the ``fully adjusted" model of \cite{step_up_primary}. The coefficients in both sets of results are similar to the main analysis.
However, both intercepts in the unadjusted analysis and the FPCA + regression intercept in the fully adjusted analysis are shifted upward compared to Figure \ref{fig:combined_int_coef_plot}, which is likely due to the choice of reference levels of the covariates.
The lack of changes in the intervention effects, though, indicates some robustness to model misspecification in the covariates.

With the exception of the first analysis with daily steps, the functional coefficients remained similar across the analyses, indicating that overall, the conclusions are not very sensitive to the preprocessing and covariates used. 
The wider confidence intervals in some analyses affect the significance of the collaborative and supportive intervention effects at the end of the intervention period, making conclusions about their effectiveness in those weeks less reliable.
However, more general conclusions, e.g. that the intervention effect is significant for the first portion of the intervention period in the collaborative and supportive arms and for the full period in the competitive arm, are interpretable and robust in our sensitivity analyses.

\section{FoSR model and penalty structure}
\label{supp:fosr_model}

\subsection{Specifying covariance structure with a functional random effect}
\label{supp:z_t}

\begin{figure}[!htb]
    \centering
    \includegraphics[width=.75\linewidth]{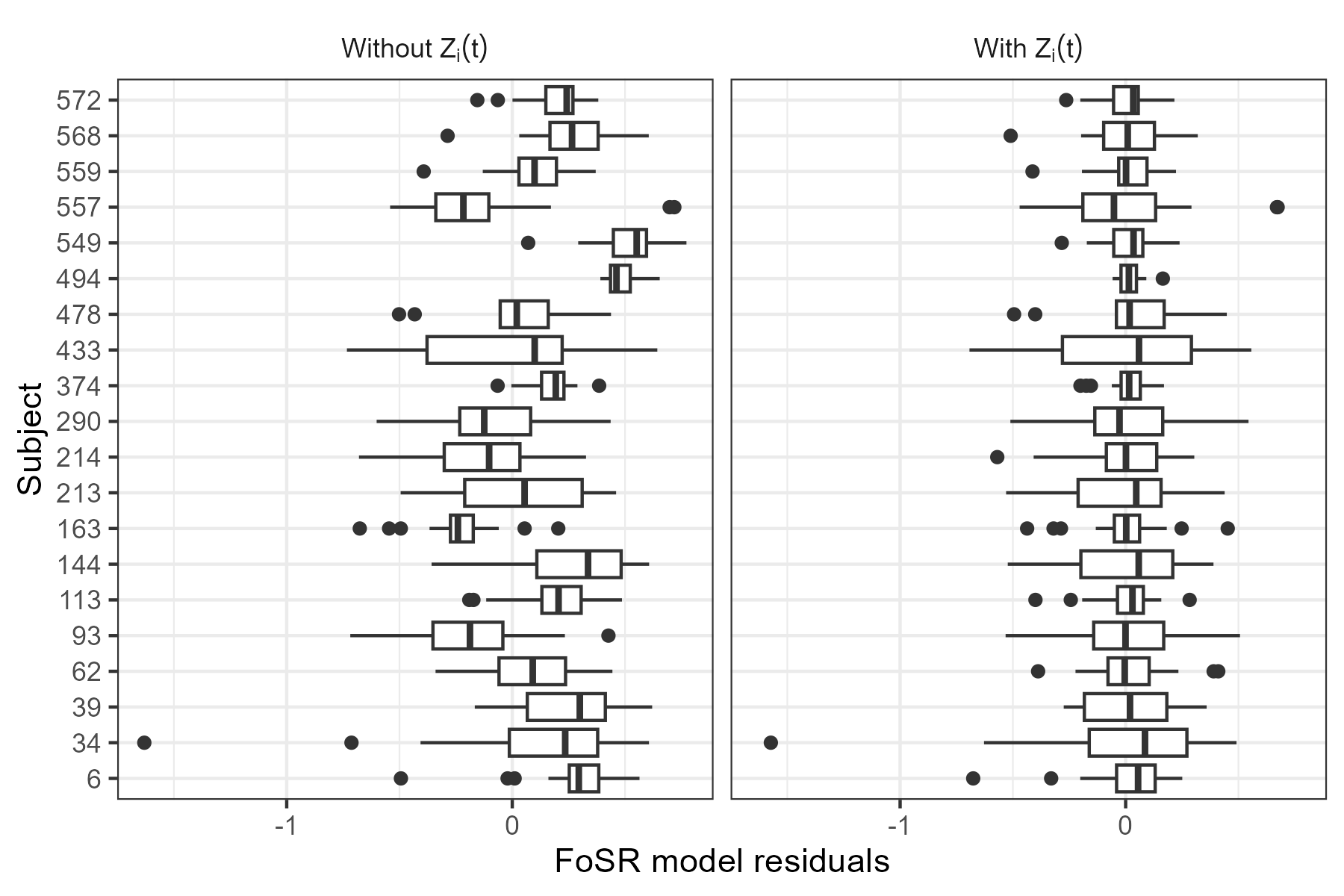}
    \caption{FoSR residuals by subject, without and with the $Z_i(t)$ random effect term. 
    The right panel contains the residuals from the FoSR model in Equation \ref{eq:fosr_eq}, and the left panel contains the residuals from the model without the subject-level effect $Z_i(t)$, which captures within-subject correlations.
    Adding the $Z_i(t)$ term centers the residuals within each subject, but there are still large differences in the variability of the residuals across subjects.
    } 
    \label{fig:residuals_boxplot}
\end{figure}

There is only one functional observation $W_i(\cdot)$ per person, so the role of the subject-level random effect $Z_i(\cdot)$ in the FoSR model Equation \ref{eq:fosr_eq} may be unclear. 
As \cite{morris2015review} explains, assuming i.i.d. residual errors is unrealistic in FDA because there is likely correlation across $t$ and large variation in subject-level deviations from the mean function.
Modeling these curve-to-curve deviations with a PC decomposition is a effective approach to capture the within-function correlation.
If enough basis functions are used for the functional random intercept, then it can flexibly capture functions for the subject-specific deviation.
In practice, only the top $K$ eigenfunctions are used for the basis, but as we discuss in Sections \ref{subsec:methods_fpca} and \ref{subsubsec:results_comparison}, the results are not sensitive to the number of eigenfunctions used.
Figure \ref{fig:residuals_boxplot} show box plots of residuals by subject for 30 subjects from an FoSR model without and with the $Z_i(t)$ term.
On the left side, without $Z_i(t)$, the means and variances of the residuals vary widely between participants. 
With $Z_i(t)$, the residuals for most subjects are centered; however, the variability of the residuals still differs by subject.

\subsection{Penalized model fitting}
\label{supp:fosr_penalty}

Penalized splines are a standard approach in the FDA literature to flexibly capture complex effects while avoiding over-fitting (see, e.g., \cite{goldsmith_penalized_fr_2011, reiss2010_fosr, aguilera_psplinesvarselection, scheipl2015famm, ivanescu2015penalized_fofr}).
A penalty makes model fit more robust to choices such as number of knots or basis functions through regularization \citep{semiparametric_regression_ruppert2003}.
\cite{luoxiao_asymp} analyzed the convergence rates of penalized splines in FDA, specifically P-splines, which we use. 
As shown in \cite{luoxiao_asymp}, the convergence rate of penalized splines can be decomposed into three components: approximation bias of the spline functions, shrinkage bias from the smoothness penalty, and variability of the penalized splines.
The approximation bias of the splines to the true function depends on the number $K$ and order $m$ of basis functions and the smoothness of the target function; it is small when the target has sufficiently many continuous derivatives and $K$ is large enough.
The shrinkage bias additionally contributed by the penalty is on the order of $\lambda^2 h_e^{-2q}$, where $\lambda$ is the smoothing parameter, $q$ is the order of the difference penalty, and $h_e = \min(K^{-1}, \lambda^{\frac{1}{2q}})$. This is of smaller order when the number of basis functions is sufficient and the penalty is not overly restrictive. 
In all our models, $m = 2$ and $q = 2$, which are standard values.
For the intervention period, we use $K = 20$ basis functions. The largest effective degrees of freedom of the functional coefficients is 4.05, which is not close to $K$, indicating that $K$ is sufficiently large.
In the follow-up period, we set $K = 7$ due to fewer time points. Still, the largest effective degrees of freedom is 2.2, so $K = 7$ is sufficient here.
In the fitted models with linear coefficients, the estimated $\lambda$ is very large, possibly leading to a slower rate of convergence by encouraging excessively smooth coefficients, though the estimation is still consistent. Given these diagnostics and the use of bootstrapping for inference, the shrinkage bias induced by the penalty is likely negligible in this case because our model parameters are fairly standard.

\section{Cohort effects in the collaborative and competitive arms}
\label{supp:interference}

\begin{figure}
    \centering
    \includegraphics[width=.75\linewidth]{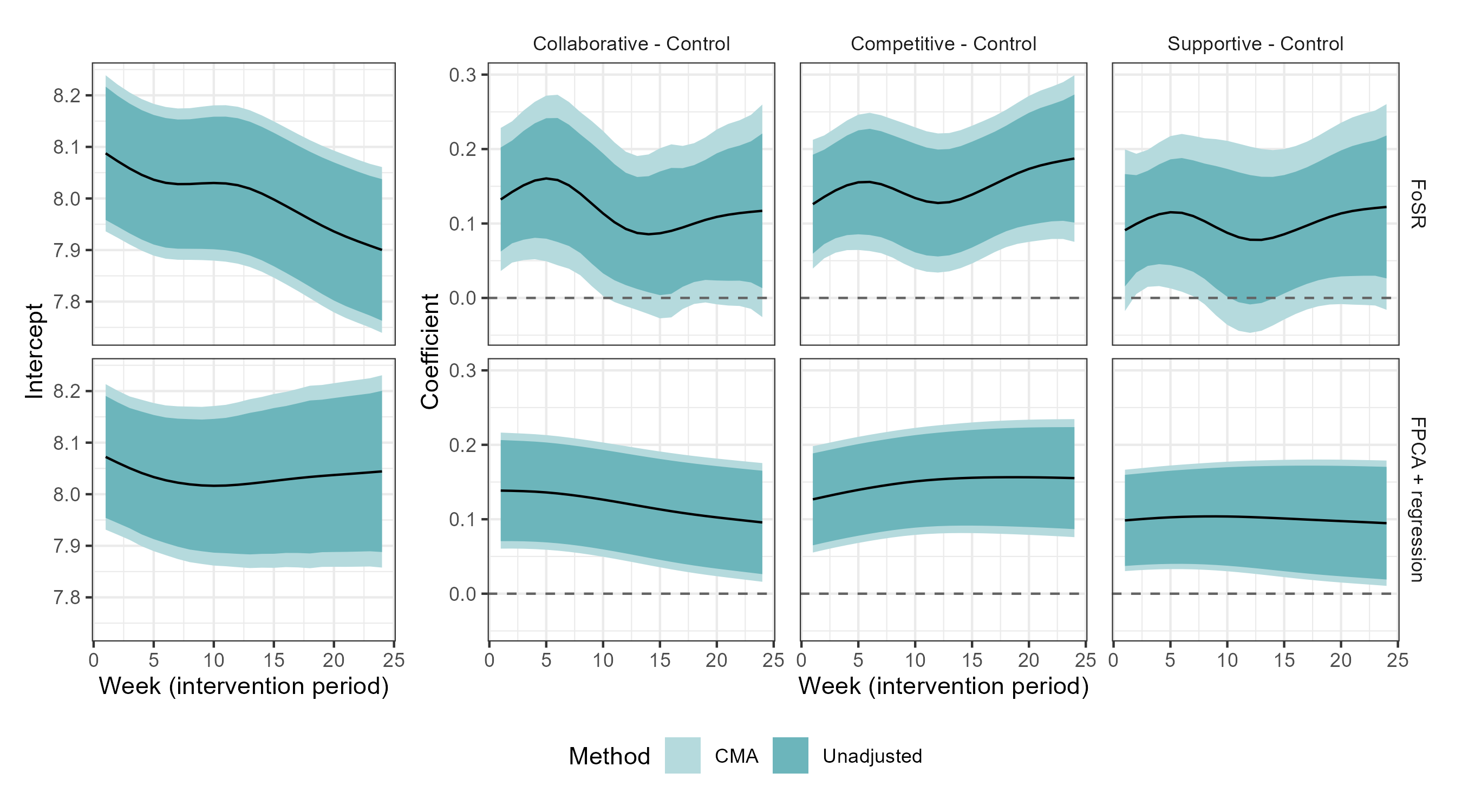}
    \caption{Functional coefficients from FoSR and FPCA + regression with unadjusted and CMA confidence intervals created from the bootstrap algorithm in Section \ref{supp:interference}.
    This is analogous to Figure \ref{fig:combined_int_coef_plot}, which uses a simple bootstrap. There are not significant qualitative differences between the two, but the confidence intervals generated by the stratified bootstrap approach are slightly larger.}
    \label{fig:s_combined_coef_interference}
\end{figure}

In the competitive and collaborative arms, participants were placed into cohorts of three and engaged with each other in the intervention.
Prior analyses of STEP UP treated all individuals as independent and did not account for cohort effects, but, in these two arms, it is possible that the outcomes of participants are correlated with others in their cohort.
To account for such correlation, we created a second bootstrap procedure for calculating the standard errors of functional coefficients, incorporating the cohort structure and randomization scheme.
Specifically, we resampled at the cohort level for the competitive and collaborative arms, and so the bootstrap variance estimates account for the inflated variance due to correlation within cohort. 
For the supportive and control arms, we resample stratifying on baseline step count: one individual is drawn, and two others in the same baseline step strata and arm are selected, until there are 151 participants selected in each arm.
The confidence intervals for the functional coefficients are given in Figure \ref{fig:s_combined_coef_interference}. 

Qualitatively, the plot coefficients and confidence intervals look similar to those in Figure \ref{fig:combined_int_coef_plot}. The confidence intervals in Figure \ref{fig:s_combined_coef_interference} are slightly larger, so there are fewer weeks in which the collaborative and supportive interventions have significant effects. Specifically, the intervention effect for the collaborative arm is significant in weeks 1-10 (was 1-11 and 18-23), and the effect for the supportive arm is significant in weeks 2-7 (was 1-8 and 20-22).
That is, incorporating the sampling and randomization schemes into the bootstrap shows that towards the end of the intervention period, the collaborative and supportive interventions do not significantly increase step count compared to the control arm.

\section{Effectiveness of intervention in follow-up period}
\label{supp:follow_up}

\begin{figure}
    \centering
    \includegraphics[width=\linewidth]{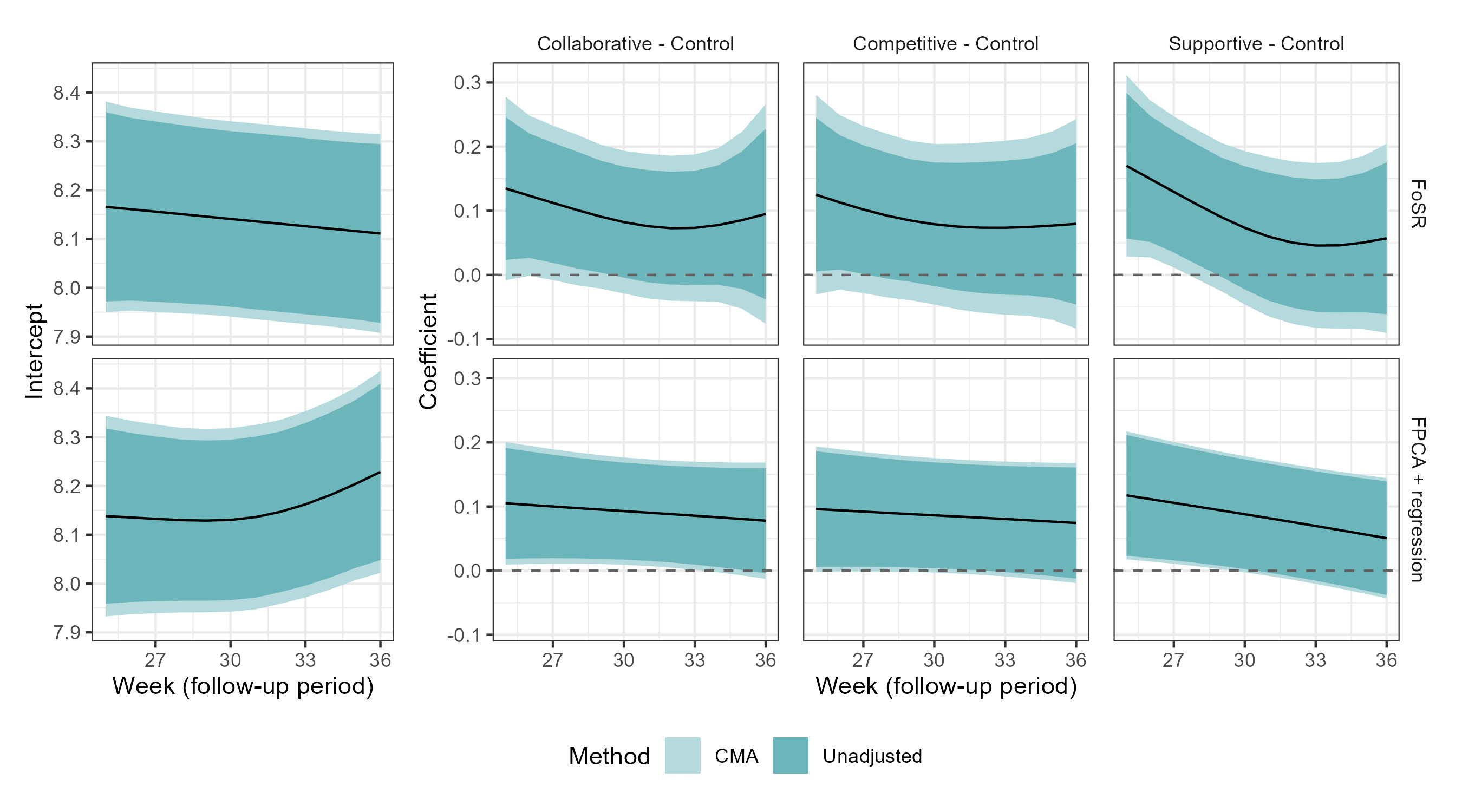}
    \caption{Functional coefficients from FoSR and FPCA + regression with unadjusted and correlation and multiplicity adjusted (CMA) confidence intervals for the follow-up period. The coefficients show effect of study arm relative to control, as in Figure \ref{fig:combined_coef_plot}. The effect at time $t$ is considered significant if the CMA CI does not contain 0. Note the outcome is log steps.}
    \label{fig:s_combined_coef_plot_post}
\end{figure}

The FoSR and FPCA + regression functional coefficients for the follow-up period are in Figure \ref{fig:s_combined_coef_plot_post}, and their contrasts are in Figure \ref{fig:s_all_contrast_plot_post}.
For FoSR, there are no CMA significant differences between any pair of arms at any time, except the supportive arm is significantly higher than control in weeks 25-27.
For FPCA + regression, the collaborative arm is significantly higher than control in weeks 25 to 33, and the supportive arm is significantly higher than control in weeks 25 to 29. 
The issue of linear FoSR coefficient estimates discussed in Subsection \ref{subsubsec:results_comparison} is present here as well. The raw coefficients for collaborative and competitive arm are linear, hence the corresponding coefficients in Figure \ref{fig:s_combined_coef_plot_post} have the same shape as the intercept.

\clearpage

\begin{table}[!htb]
\centering
\begin{tabular}{llrrll}
  \hline
Eigenfunction & Term & Estimate & Std. error & P-value & Adj. p-value \\ 
  \hline
Eigenfunction 1 & Age & 0.01 & 0.01 & 0.016 & 0.205 \\ 
   & Male & 0.09 & 0.12 & 0.462 & 1 \\ 
   & Baseline mean daily steps & 0.00 & 0.00 & $<$.001 & $<$.001 \\ 
   & Previously used a wearable & -0.03 & 0.11 & 0.786 & 1 \\ 
   & PSQI overall & -0.04 & 0.12 & 0.709 & 1 \\ 
   & Extroversion & 0.01 & 0.07 & 0.894 & 1 \\ 
   & Agreeableness & -0.24 & 0.11 & 0.027 & 0.323 \\ 
   & Conscientiousness & 0.27 & 0.12 & 0.028 & 0.323 \\ 
   & Neuroticism & -0.01 & 0.09 & 0.886 & 1 \\ 
   & Openness & 0.02 & 0.10 & 0.816 & 1 \\ 
   & Grit & -0.25 & 0.13 & 0.058 & 0.576 \\ 
   & ESE: sticking to it & 0.30 & 0.08 & $<$.001 & 0.004 \\ 
   & DOSPERT: health/safety & 0.08 & 0.06 & 0.181 & 1 \\ 
   & DOSPERT: social & -0.10 & 0.06 & 0.1 & 0.898 \\ 
   & MOS SS overall & -0.00 & 0.00 & 0.992 & 1 \\ 
   & Start day & -0.01 & 0.00 & $<$.001 & 0.004 \\ 
   & Collaborative & 0.54 & 0.15 & $<$.001 & 0.004 \\ 
   & Competitive & 0.71 & 0.15 & $<$.001 & $<$.001 \\ 
   & Supportive & 0.49 & 0.15 & 0.001 & 0.018 \\ 
  Eigenfunction 2 & Age & -0.00 & 0.00 & 0.19 & 1 \\ 
   & Male & 0.02 & 0.04 & 0.598 & 1 \\ 
   & Baseline mean daily steps & -0.00 & 0.00 & $<$.001 & $<$.001 \\ 
   & Previously used a wearable & -0.01 & 0.03 & 0.656 & 1 \\ 
   & PSQI overall & 0.04 & 0.04 & 0.293 & 1 \\ 
   & Extroversion & -0.02 & 0.02 & 0.396 & 1 \\ 
   & Agreeableness & -0.06 & 0.03 & 0.085 & 1 \\ 
   & Conscientiousness & -0.00 & 0.04 & 0.965 & 1 \\ 
   & Neuroticism & -0.01 & 0.03 & 0.736 & 1 \\ 
   & Openness & -0.01 & 0.03 & 0.658 & 1 \\ 
   & Grit & 0.01 & 0.04 & 0.839 & 1 \\ 
   & ESE: sticking to it & 0.00 & 0.02 & 0.972 & 1 \\ 
   & DOSPERT: health/safety & 0.01 & 0.02 & 0.503 & 1 \\ 
   & DOSPERT: social & 0.01 & 0.02 & 0.425 & 1 \\ 
   & MOS SS overall & 0.00 & 0.00 & 0.576 & 1 \\ 
   & Start day & -0.00 & 0.00 & 0.09 & 1 \\ 
   & Collaborative & -0.09 & 0.05 & 0.051 & 0.964 \\ 
   & Competitive & 0.02 & 0.05 & 0.619 & 1 \\ 
   & Supportive & -0.02 & 0.05 & 0.733 & 1 \\ 
   \hline
\end{tabular}
\caption{Full results of linear regression in FPCA + regression in the intervention period. Adjusted p-values are calculated using a Holm-Bonferroni correction \citep{holm_bonferroni} to control family-wise error rate for each model. Full descriptions of the survey measures can be found in \citep{step_up_latent}.}
\label{table:inteffull}
\end{table}

\begin{table}[!htb]
\centering
\begin{tabular}{llrrll}
  \hline
Eigenfunction & Term & Estimate & Std. error & P-value & Adj. p-value \\ 
  \hline
Eigenfunction 1 & Age & -0.00 & 0.01 & 0.517 & 1 \\ 
   & Male & -0.04 & 0.12 & 0.707 & 1 \\ 
   & Baseline mean daily steps & -0.00 & 0.00 & $<$.001 & $<$.001 \\ 
   & Previously used a wearable & -0.13 & 0.11 & 0.21 & 1 \\ 
   & PSQI overall & 0.13 & 0.11 & 0.258 & 1 \\ 
   & Extroversion & 0.00 & 0.07 & 0.96 & 1 \\ 
   & Agreeableness & 0.30 & 0.11 & 0.005 & 0.097 \\ 
   & Conscientiousness & 0.04 & 0.12 & 0.746 & 1 \\ 
   & Neuroticism & 0.02 & 0.08 & 0.843 & 1 \\ 
   & Openness & 0.09 & 0.10 & 0.345 & 1 \\ 
   & Grit & 0.06 & 0.13 & 0.64 & 1 \\ 
   & ESE: sticking to it & -0.18 & 0.08 & 0.026 & 0.387 \\ 
   & DOSPERT: health/safety & -0.06 & 0.06 & 0.278 & 1 \\ 
   & DOSPERT: social & -0.04 & 0.06 & 0.498 & 1 \\ 
   & MOS SS overall & -0.00 & 0.00 & 0.826 & 1 \\ 
   & Start day & 0.01 & 0.00 & $<$.001 & $<$.001 \\ 
   & Collaborative & -0.33 & 0.15 & 0.026 & 0.387 \\ 
   & Competitive & -0.36 & 0.14 & 0.012 & 0.204 \\ 
   & Supportive & -0.36 & 0.15 & 0.014 & 0.223 \\ 
  Eigenfunction 2 & Age & -0.00 & 0.00 & 0.039 & 0.788 \\ 
   & Male & -0.03 & 0.03 & 0.371 & 1 \\ 
   & Baseline mean daily steps & -0.00 & 0.00 & 0.256 & 1 \\ 
   & Previously used a wearable & -0.01 & 0.03 & 0.738 & 1 \\ 
   & PSQI overall & 0.03 & 0.03 & 0.315 & 1 \\ 
   & Extroversion & 0.01 & 0.02 & 0.597 & 1 \\ 
   & Agreeableness & -0.00 & 0.03 & 0.911 & 1 \\ 
   & Conscientiousness & 0.01 & 0.03 & 0.85 & 1 \\ 
   & Neuroticism & -0.01 & 0.02 & 0.593 & 1 \\ 
   & Openness & 0.03 & 0.02 & 0.262 & 1 \\ 
   & Grit & -0.00 & 0.03 & 0.882 & 1 \\ 
   & ESE: sticking to it & 0.01 & 0.02 & 0.682 & 1 \\ 
   & DOSPERT: health/safety & -0.00 & 0.01 & 0.797 & 1 \\ 
   & DOSPERT: social & 0.01 & 0.01 & 0.722 & 1 \\ 
   & MOS SS overall & -0.00 & 0.00 & 0.997 & 1 \\ 
   & Start day & -0.00 & 0.00 & 0.867 & 1 \\ 
   & Collaborative & -0.03 & 0.04 & 0.484 & 1 \\ 
   & Competitive & -0.02 & 0.03 & 0.576 & 1 \\ 
   & Supportive & -0.07 & 0.03 & 0.041 & 0.788 \\ 
   \hline
\end{tabular}
\label{table:posteffull}
\caption{Full results of linear regression in FPCA + regression in the follow-up period. Full descriptions of the survey measures can be found in \citep{step_up_latent}.}
\end{table}

\clearpage

\begin{figure}
    \centering
    \includegraphics[width=0.75\linewidth]{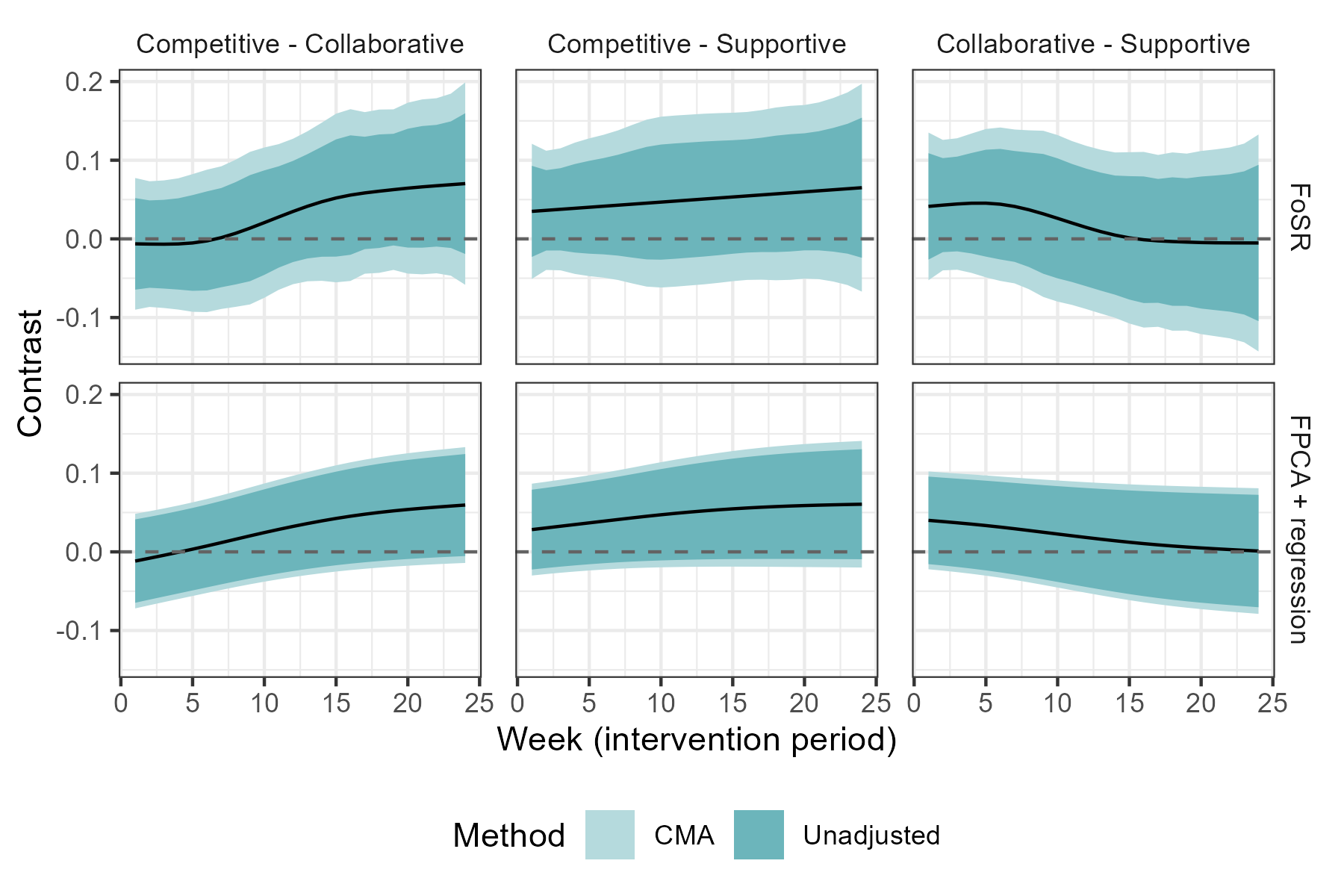}
    \caption{Contrasts of functional coefficients from FoSR and FPCA + regression with unadjusted correlation and multiplicity adjusted adjusted (CMA) confidence intervals.}
    \label{fig:s_combined_contrast_plot}
\end{figure}

\begin{figure}[!htb]
    \centering
    \includegraphics[width=0.75\linewidth]{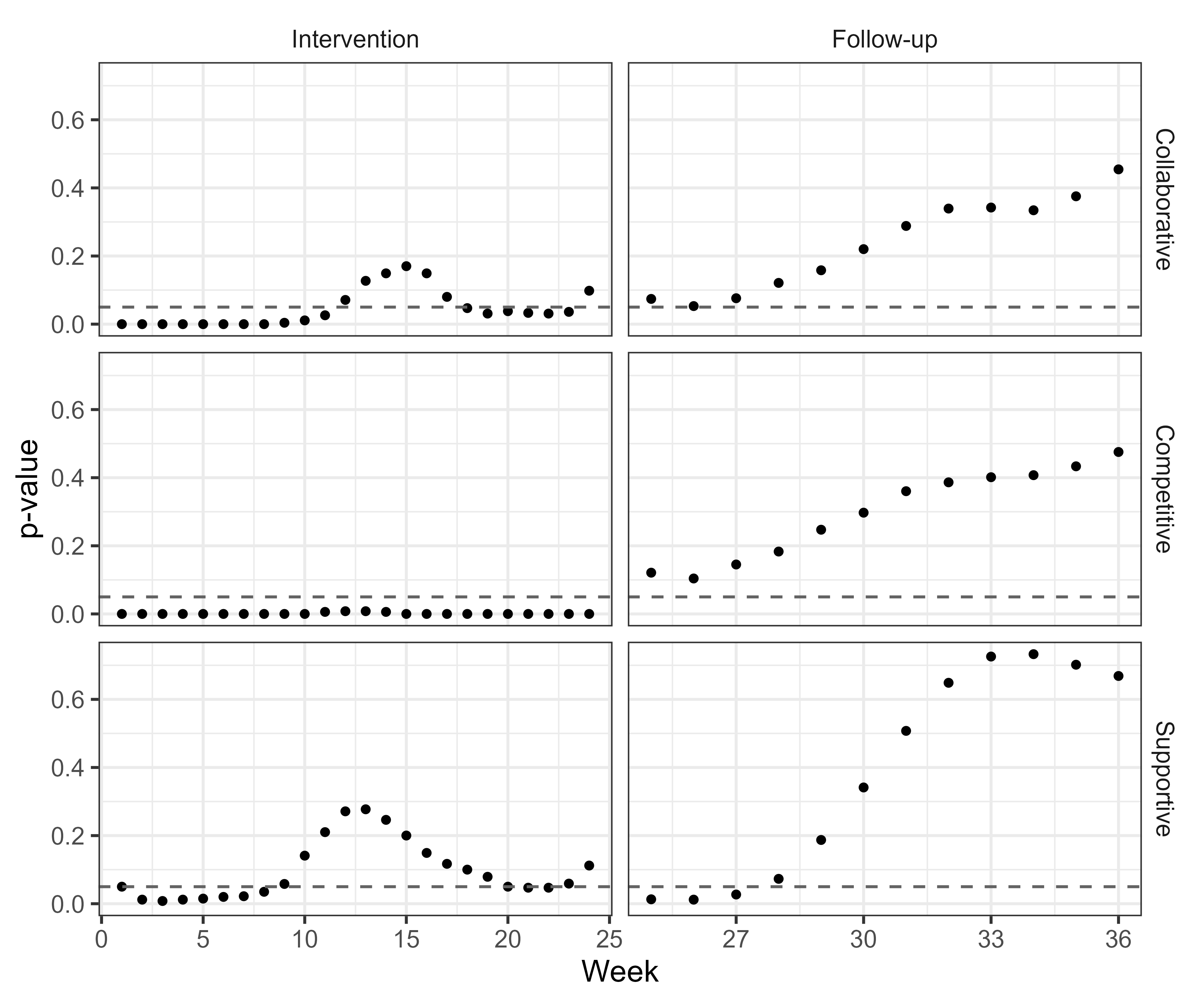}
    \caption{Correlation and multiplicity adjusted (CMA) pointwise $p$-values from the FoSR models for testing the null hypothesis that the coefficient function is 0.}
    \label{fig:s_fosr_pvalue_plot}
\end{figure}

\begin{figure}
    \centering
    \includegraphics[width=.8\linewidth]{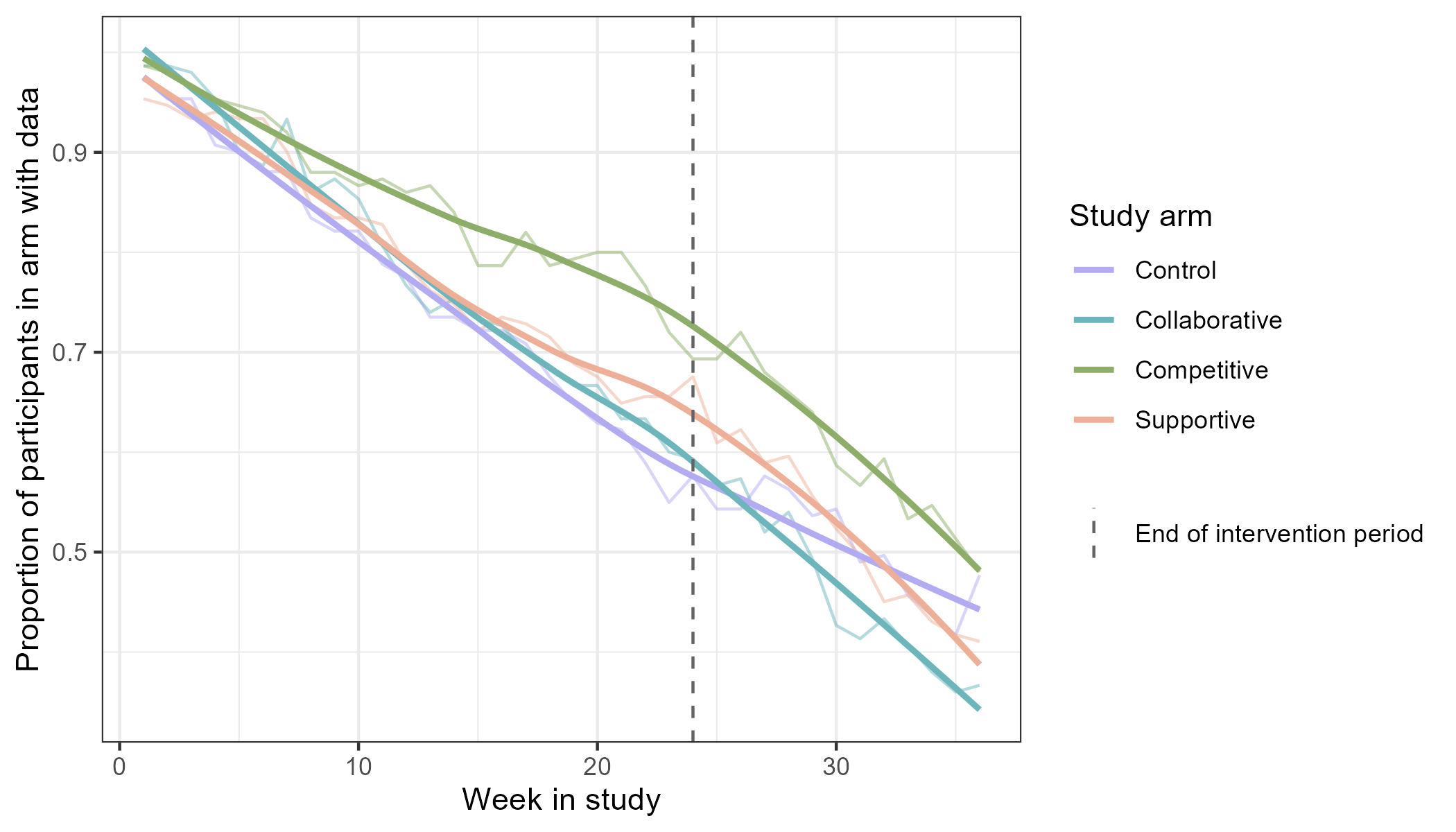}
    \caption{Missingness by study arm. Missingness increased steadily as the study progressed, and also varied by arm. The competitive arm had less missingness than the other arms, particularly in the middle of the study}
    \label{fig:s_miss_by_treat}
\end{figure}

\begin{figure}
    \centering
    \includegraphics[width=.75\linewidth]{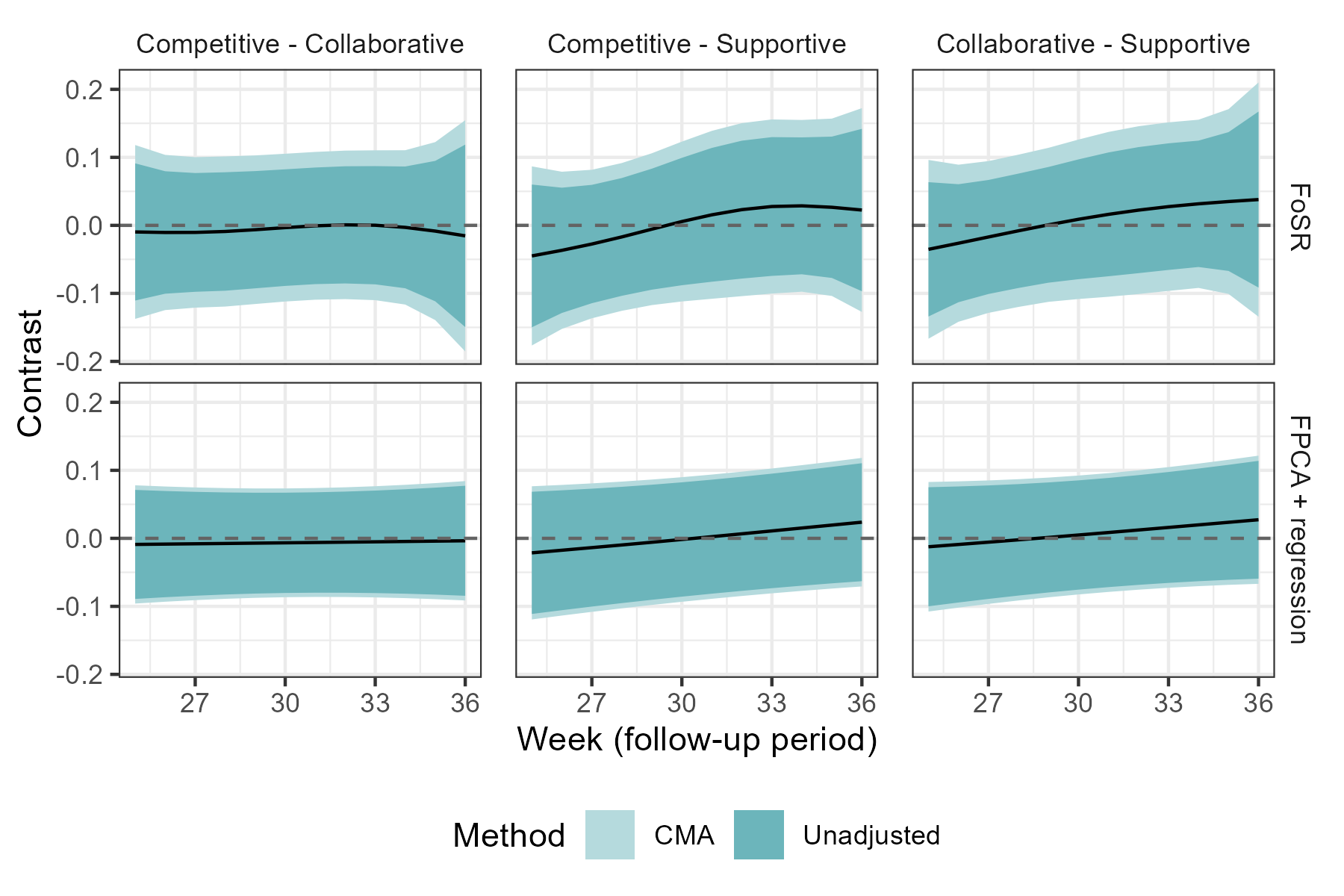}
    \caption{Contrasts of functional coefficients from FoSR and FPCA + regression with unadjusted correlation and multiplicity adjusted adjusted (CMA) confidence intervals for the follow-up period. This is analogous to Figure \ref{fig:s_combined_contrast_plot}.}
    \label{fig:s_all_contrast_plot_post}
\end{figure}

\end{document}